\documentclass[12pt]{report}

\usepackage[utf8]{inputenc}
\usepackage[english]{babel}
\usepackage[letterpaper, margin=30mm]{geometry}
\usepackage[parfill]{parskip}
\usepackage{csquotes}
\usepackage[natbib,uniquelist=false,backend=biber,style=authoryear-comp,maxbibnames=99,maxcitenames=2]{biblatex}
\usepackage{graphicx}
\usepackage{amsmath}
\usepackage{mathtools}
\DeclareMathOperator*{\argmin}{argmin}
\usepackage{amsfonts} % \mathbb{R}
\usepackage{array}
\usepackage{braket}
\usepackage{xcolor}
\usepackage[colorlinks=true,linkcolor=blue,citecolor=blue,urlcolor=blue,linktoc=all]{hyperref}
\usepackage{abstract}
\graphicspath{{Figuras/}}
\newrobustcmd*{\parentexttrack}[1]{%
  \begingroup
  \blx@blxinit
  \blx@setsfcodes
  \blx@bibopenparen#1\blx@bibcloseparen
  \endgroup}

\AtEveryCite{%
  }
  
\let\oldlistoffigures\listoffigures
\makeatletter
\renewcommand{\listoffigures}{%
  \patchcmd{\@makeschapterhead}{\vspace*{50\p@}}{\relax}{}{}%
  \oldlistoffigures%
  \patchcmd{\@makeschapterhead}{\relax}{\vspace*{50\p@}}{}{}%
}  
\makeatother

\begin{document}

\newcommand{\TODO}[1]{\textcolor{red}{#1}}

\newcommand{\titulo}{Motion Matching for Character Animation and Virtual Reality Avatars in Unity}
\newcommand{\nombreestudiante}{Jose Luis Ponton}
\newcommand{\nombredirector}{Nuria Pelechano}
\newcommand{\nombrecodirector}{Carlos Andújar}
\newcommand{\fecha}{June 27th, 2022}  % Definir solo el año de presentación

\renewcommand{\listtablename}{Índice de tablas} 
\renewcommand{\tablename}{Tabla} 

\begin{titlepage}
	\centering
	\includegraphics[width=135mm]{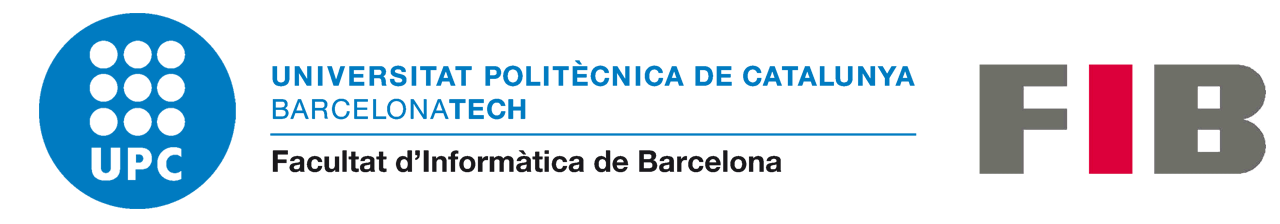}\par
	\vspace{1cm}
    {\large
    Universitat Politècnica de Catalunya (UPC) - BarcelonaTech\par
    Facultat d'Informàtica de Barcelona (FIB)\par
    Master in Innovation and Research in Informatics \\
    Computer Graphics and Virtual Reality\par}
    \vspace{1cm}
	{\large Master Thesis\par}
	\vspace{1cm}
	{\LARGE\bfseries \titulo \par}
	{\large
	\vfill
	\nombreestudiante\par
	\vfill
	\emph{Supervisors:}\par\vspace{2mm}
    \nombredirector \par\vspace{2mm}
    \nombrecodirector
    \vfill}
	{\large\fecha\par}
\end{titlepage}

\renewcommand{\abstractnamefont}{\normalfont\LARGE\bfseries}
\begin{abstract}
\vspace{0.5cm}
    Real-time animation of virtual characters has traditionally been accomplished by playing short sequences of animations structured in the form of a graph. These methods are time-consuming to set up and scale poorly with the number of motions required in modern virtual environments. The ever-increasing need for highly-realistic virtual characters in fields such as entertainment, virtual reality, or the metaverse has led to significant advances in the field of data-driven character animation. Techniques like Motion Matching have provided enough versatility to conveniently animate virtual characters using a selection of features from an animation database. Data-driven methods retain the quality of the captured animations, thus delivering smoother and more natural-looking animations. In this work, we researched and developed a Motion Matching technique for the Unity game engine. In this thesis, we present our findings on how to implement an animation system based on Motion Matching. We also introduce a novel method combining body orientation prediction with Motion Matching to animate avatars for consumer-grade virtual reality systems.
\end{abstract}

\renewcommand{\abstractname}{Acknowledgements}
\begin{abstract}
\vspace{0.5cm}
I would like to express my gratitude to my research supervisors, Nuria and Carlos, for their help, dedication and patience throughout the development of this work. Not only they have provided continuous valuable advice, but they also motivated me to get over the difficulties.

I also want to thank my friends for their constant support and helping me when things were tough during my studies. Special thanks to Haoran, with whom I shared countless hours of motion capture and VR development.

Finally, I want to thank my family for encouraging me to continue researching and for all the support that kept me motivated during my master's degree studies.
\end{abstract}

\def\table{\def\figurename{Table}\figure} % Table of Figures and Tables together
\let\endtable\endfigure 
{
    \renewcommand\listfigurename{List of Figures and Tables}

    \hypersetup{linkcolor=black}
    
    \tableofcontents
    \listoffigures
%    \listoftables
}

\chapter{Introduction} \label{chap:intro}

\section{Motivation}
The recent pandemic has forced most of the world to work remotely using video conferencing, which has proven to provide substantial benefits in terms of saving travel expenses and facilitating attendance. While video conferencing allows us to see each other and facilitates non-verbal communication with respect to audio conferencing, interaction is essentially 2D and thus a limiting factor. A widely shared vision for real collaboration is based on Virtual Reality (VR), where several users wearing headsets should be able to interact with each other freely. One-on-one meetings and small group discussions might highly benefit from having such enhanced communication through facial expressions, gestures, pointing and other real-life interactions. As its video conferencing counterpart, VR-based remote collaboration removes physical barriers and can connect team members from around the world with diverse profiles, e.g., in terms of age, gender, race, culture, social status and functional abilities.

The most common solution to enable non-verbal communication in VR is to use avatars that represent, either in a minimalist or detailed way, our body and that of the collaborators. Avatars require obtaining real-time data that allow placing, orienting, and reproducing the users’ gestures.

\begin{figure}[h]
  \centering
  \includegraphics[width=1.0\linewidth]{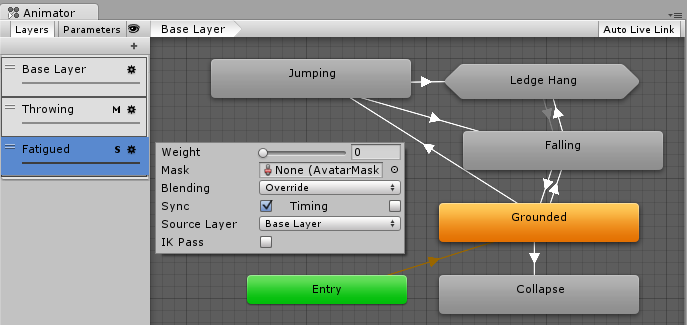}
  \caption[Animation Graph]{\label{fig:intr:animation_controller} Graph containing animations clips and their transitions for a virtual character. It contains several layers depending on the character's action. Source: \href{https://docs.unity3d.com}{https://docs.unity3d.com}}
\end{figure}

Traditionally, character animation is accomplished by playing short sequences (animation clips) of animations structured in the form of a graph (see Figure \ref{fig:intr:animation_controller}). Creating these animations, the graph and the transitions is an time-consuming task and may not produce high-quality animations due to the need to define all possible movements and transitions between poses. The increasing interest in obtaining high-quality animations has led to the development of new data-driven methods such as Motion Matching \citep{clavet2016}. In these methods, animations are automatically extracted from an animation database according to a feature vector defined by the user’s input and the current pose of the virtual character.

On the one hand, Motion Matching has become a crucial algorithm behind some of the recent AAA game productions such as \emph{For Honor} by \emph{Ubisoft} and \emph{The Last of Us 2} by \emph{Naughty Dog}. On the other hand, the high-dimensionality of VR input (3 trackers with 6 DOF each vs. 1 joystick with 2 DOF) and the unpredictability of the user’s movement have become a challenge for developing data-driven animation methods for VR avatars.

Although high-quality results can be obtained using data-driven character animation, still, some limitations exist: high-memory requirements, lack of a general method for conveniently representing any movement, need for expensive equipment to generate the animation database, and applicability to different-sized characters, among others. Therefore, research in data-driven character animation is necessary to overcome the current limitations. In this work, we research and implement a Motion Matching method for Unity and present a novel method for animating VR avatars. We combine our custom Motion Matching implementation with a neural network to animate a full-body VR avatar using only the Head-Mounted Display (HMD) and the two controllers commonly available in all consumer-grade VR systems.

\section{Objectives}

The main objective of this work is to develop and research data-driven methods for animating virtual characters with an emphasis on VR avatars. Specifically, we focus our research and development on the Motion Matching algorithm. Therefore, the main objective can be divided into the following sub-objectives:

\begin{itemize}
    \item Implementation of Motion Matching in Unity. This sub-objective requires studying and researching current systems and research papers using Motion Matching. We decided to implement the system in Unity to increase its reusability in multiple projects and facilitate its integration with VR systems.
    \item Research and implementation of a novel system to generate high-quality animations for VR avatars using data-driven methods.
    \item Use a motion capture system (Xsens) to capture animation databases suitable for Motion Matching and VR avatars.
\end{itemize}

Although Motion Matching has been implemented in some companies’ in-house game engines and has been presented in conferences \citep{buttner2015, clavet2016}, with this work, we want to provide a self-contained document formulating and presenting all algorithms and methods involved for the implementation of Motion Matching.

In the field of VR, expensive motion-capture suits exist to animate full-body avatars. However, consumer-grade VR systems are only equipped with one HMD and two controllers. Despite the new advances in the animation of virtual characters, traditional techniques are still used in VR. We believe data-driven animations for VR are the necessary step to tackle the increasing requirements for high-quality animations in mainstream VR devices. We hope this work will provide a first foundation for using data-driven methods for VR avatars.

\newpage
\section{Document Structure}
The overall structure of this work takes the form of six chapters and is divided into two major topics: Motion Matching and VR avatars. Chapters \ref{chap:intro}, \ref{chap:back}, \ref{chap:results} and \ref{chap:conclusions} present general concepts applicable to both topics, while Chapter~\ref{chap:mm} and \ref{chap:vr} specifically address Motion Matching and VR avatars, respectively.

A brief description of each Chapter:
\begin{itemize}
    \item \textbf{Chapter~\ref{chap:intro} Introduction} states the motivation and objectives of this work.
    \item \textbf{Chapter~\ref{chap:back} Background} introduces the state-of-the-art for data-driven character animation and VR avatars. Some preliminary definitions are defined for their use throughout this work.
    \item \textbf{Chapter~\ref{chap:mm} Animation System Implementation} formulates and describes our proposed implementation for the animation system based on Motion Matching. This chapter focuses on the animation of virtual characters in general. The concepts described here can be used for desktop and VR applications.
    \item \textbf{Chapter~\ref{chap:vr} Virtual Reality Avatars} explicitly addresses the problem of animating avatars in VR. This chapter describes how to use Motion Matching for VR avatars and how to predict the body orientation using Deep Learning. 
    \item \textbf{Chapter~\ref{chap:results} Discussion and Results} contains images showing the resulting animations of this work. It also includes discussions comparing our method with the standard approaches and evaluates the system.
    \item \textbf{Chapter~\ref{chap:conclusions} Conclusions and Future Work} gives a summary of this work, and areas for further research are identified.
\end{itemize}
\chapter{Background} \label{chap:back}

\section{Related Work}

\subsection{Data-driven character animation}
Over the past 20 years, the animation community has developed multiple data-driven algorithms. The first approaches used graphs to provide an abstract representation of motion sequences and plausible transitions between them. Although this representation can provide computational advantages due to the inherently discrete structure, creating graphs that allow for responsive user control is challenging since transitions happen only in predetermined places. Also, only poses in the graph can be represented, making tasks such as physically-based characters difficult. Numerous works \citep{kovar2002, arikan2002, lee2002} explored the idea of formulating motion synthesis as a search over a graph where poses are nodes and similar poses are connected, thus allowing the use of a single animation database and automatically extracting sequences of plausible poses.

Other works tried to minimize the limitations of pure graph-based representations. For instance, trying to augmented the graph's connectivity using parametric motion graphs \citep{heck2007, shin2006} or by increasing the number of transitions \citep{arikan2005, yin2005}. \citet{lee2010} extended the idea of motion graph to a continuous space by using reinforcement learning to interpolate the nearest neighbors and synthesize an animation to accomplish the desired goals.

In order to avoid the automatic graph construction, other research uses linear methods such as Principal Component Analysis (PCA) to synthesize animation from low-dimensional features \citep{chai2005}, or non-linear kernel methods \citep{levine2012}. Recently, neural networks have been used to synthesize locomotion animations for humans successfully \citep{holden2017} and quadruped animals such as dogs \citep{zhang2018}. However, neural networks can be challenging to train and control. Moreover, high-frequency details may be smoothed.

\subsection{Avatars in virtual reality}

The importance of self-avatars has long been recognized from various perspectives, including user performance, distance perception, cognitive load, Sense of Embodiment (SoE) and presence. With the popularity of HMD-based VR devices, the impact of the avatar's visual fidelity has drawn much attention. Due to the limited tracking information provided by consumer-grade VR, the floating hands representation has become the most common form of self-avatars in VR applications and games. However, recent studies have shown that hands-only representations provide little SoE. \citet{JungandHughes2016} conducted a study with hand-focused tasks to investigate the effect of inferred body parts on the Sense of Body Ownership (SoBO), one of the subcomponents of SoE. Their results suggest that the inferred lower body leads to a higher level of SoBO than the no lower body condition.

\citet{fribourg2020} investigated the effect of self-avatars' appearance, control and point-of-view on SoE. They found that most users were not satisfied with abstract self-avatars (e.g., five spheres for body extremities) when performing tasks like yoga, walking and kicking. \citet{galvandebarba2020} explored the effect of different levels of body parts' animation on Plausibility Illusion and Sense of Control (SoC). They concluded that adding foot tracking to full-body self-avatars increased the SoC of the users the most.  

Additionally, full-body self-avatars and hand-only avatars lead to different behaviors and cognitive loads in VR. \citet{pan2019a} demonstrated that a full-body avatar could reduce users' cognitive load when performing spatial reasoning tasks, in contrast to hand-only avatars. \citet{ogawa2020} suggested that realistic full-body self-avatars discouraged people more effectively from walking through virtual walls than hand-only representations.

\subsection{Data-driven VR avatars}

%\haoran{paragraph summary: our problem = sparse input for reconstruct human pose/motion + real-time constrain.}
Controlling a self-avatar in VR is equivalent to using sparse high-level input to reconstruct the human pose and motion with real-time constraints. \citet{Ellis2004} demonstrated that less than 16\;ms end-to-end latency is necessary to achieve perceptual stability in virtual environments. On top of the highly under-constrained nature of pose reconstruction, achieving low latency makes self-avatar control in VR a challenging problem.

Various user input data from consumer-grade devices can be used to synthesize real-time full-body character animation. For example, some studies used optical data from egocentric cameras mounted in a baseball cap \citep{xu2019mo}, a HMD \citep{tome2020,yang2022}, controllers \citep{Ahuja2022}, or glasses \citep{zhao2021} to estimate the body pose. Egocentric cameras suffer from extreme perspective distortion and self-occlusion, leading to inadequate tracking information for the lower body. Recent studies show that a sparse set of Inertial Measurement Units (IMUs) could accurately reconstruct full-body human motion with an accuracy similar to that of commercial IMU suits. For instance, DIP \citep{huang2018} and TransPose \citep{yi2021} use learning-based methods to accurately reconstruct the full-body pose with 6 IMUs mounted on users' wrists, knees, head and pelvis. The latency of state-of-the-art solutions based on optical data or IMUs remains high for VR applications.

Several studies propose real-time methods for reconstructing the pose from consumer-grade VR devices, including HMD, controllers, and trackers. The head, pelvis, hands and feet are tracked in a six-point tracking setting. The number of tracked points varies from three to six. IK solvers can calculate the rotations of untracked joints and reconstruct the full-body pose provided enough information about end-effectors is available \citep{FinalIK, Ponton2022}. A more detailed discussion about IK methods for reconstructing the human body in VR can be found in the survey by \citet{casermanSurvey2020}.

Four-point tracking commonly includes HMD, controllers, and an additional tracker on the pelvis \citep{yang2021} or ankles \citep{caserman2019}. This configuration eliminates the infrared occlusion and foot-floor impact problems with feet trackers. \citet{yang2021} propose a velocity-based recurrent neural network (RNN) model that accurately predicts low-body pose in real-time and could generalize to different body shapes. However, it still needs an additional tracker on the pelvis and post-processing to eliminate foot sliding. Their 45 fps frame rate is not enough for real-time VR applications.

Three-point tracking, i.e., only HMD and controllers, has been studied to obtain full-body poses in VR. Learning-based methods like variational autoencoders and Recurrent Neural Networks (RNN) models \citep{dittadi2021} can generate full-body poses from the three-point tracking data; however, while these methods replicate accurate motions for the upper body, but not for the lower body because of the lack of training data with various leg movements or lack of tracking information for the feet. In our work, the locomotion data also covers walking, squatting, and running. We aim to provide realistic lower-body motion instead of replicating user leg movements.

CoolMove \citep{ahuja2021} uses $k$-Nearest-Neighbors ($k$-NN) to generate full-body animations in VR, including boxing, basketball, climbing, running and swimming. They extract features from both the motion database and the live input, along with the position and orientation of HMD and controllers. Matched motion candidates returned by $k$-NN are blended with proportional weights to form the output pose. Their result demonstrates that the generated poses could lead to higher SoA when observed from a third-person view but lower embodiment than the IK solution due to the positional error between users' input and generated poses.

\section{Preliminary Definitions}\label{sec:b:pre}
This section presents some concepts that will be used throughout this work. It includes definitions of computer animation concepts and mathematical notation used in this document.

\subsection{Rotations}

To represent rotations, we usually use quaternions, an alternative to the conventional axis-angle representation. It stores the same information but in a more mathematically convenient form. One important property is that it can easily be interpolated, and a series of quaternions can be concatenated to create a single representation. A quaternion is a four-tuple of real numbers $(w,x,y,z)$ or equivalent $(w, \mathbf{v})$, consisting of a scalar $w$ and a three-dimensional vector $\mathbf{v}$.

The reader is referred to the book Computer Animation by \citet{parent2012} for a detailed explanation of quaternions. Here, we summarize the main equations needed for this work. Quaternion multiplication is associative but not commutative, and it is defined as follows:
\begin{align}
    (w, \mathbf{v}) (w', \mathbf{v'}) = (w w' - \mathbf{v} \cdot \mathbf{v'}, w \mathbf{v'} + w' \mathbf{v} + \mathbf{v} \times \mathbf{v'})
\end{align}
The inverse of a quaternion $\mathbf{q}^{-1} = (w, \mathbf{v})^{-1}$ is obtained by negating its vector part and dividing both parts by the magnitude squared:
\begin{align}
    \mathbf{q}^{-1} = (1/\|\mathbf{q}\|)^2 (w, -\mathbf{v}) &\qquad \text{where} \; \|\mathbf{q}\| = \sqrt{w^2 + x^2 + y^2 + z^2}
\end{align}
To rotate a vector $\mathbf{u} \in \mathbb{R}^3$ using a quaternion $\mathbf{q}$ we represent the vector as $(0, \mathbf{u})$:
\begin{align} \label{eq:b:rot}
 Rot_{\mathbf{q}}(\mathbf{u}) = \mathbf{q} (0, \mathbf{u}) \mathbf{q}^{-1}
\end{align}
We define a rotation matrix $\mathbf{R}$ with columns $(\mathbf{x}, \mathbf{y}, \mathbf{z})$:
\begin{align} \label{eq:b:mat_from_columns}
    \mathbf{R}(\mathbf{x}, \mathbf{y}, \mathbf{z}) = 
    \begin{pmatrix}
        x_0 & y_0 & z_0 \\
        x_1 & y_1 & z_1 \\
        x_2 & y_2 & z_2
    \end{pmatrix}
\end{align}
A rotation matrix $\mathbf{R}$ can be converted into a quaternion with the following equations:
\begin{align} \label{eq:b:mat_to_quat}
\begin{split}
 Rot(\mathbf{R}) &= (w, x, y, z) \\
 x &= \frac{\sqrt{\mathbf{R}_{0,0} + 1 - 2 w^2}}{2} \\
 y &= \frac{\sqrt{\mathbf{R}_{1,1} + 1 - 2 w^2}}{2} \\
 z &= \frac{\sqrt{\mathbf{R}_{2,2} + 1 - 2 w^2}}{2} \\
 w &= \frac{\sqrt{\mathbf{R}_{0,0} + \mathbf{R}_{1,1} + \mathbf{R}_{2,2} + 1}}{2}
\end{split}
\end{align}
A unit quaternion ($\mathbf{q} = (w, x, y, z)\ |
     \ w^2 + x^2 + y^2 + z^2 = 1$) can be also converted back to a rotation matrix $\mathbf{R}$:
\begin{align} \label{eq:b:quat_to_mat}
    \mathbf{R}(\mathbf{q}) =
    \begin{pmatrix*}[l]
        1 - 2y^2 - 2z^2           & \phantom{1 - {}}2xy - 2wz & \phantom{1 - {}}2xz + 2wy \\
        \phantom{1 - {}}2xy + 2wz & 1 - 2x^2 - 2z^2           & \phantom{1 - {}}2yz - 2wx \\
        \phantom{1 - {}}2xy - 2wy & \phantom{1 - {}}2yz + 2wx & 1 - 2x^2 - 2y^2 \\
    \end{pmatrix*}
\end{align}
To create a quaternion view rotation given a unit length forward $\mathbf{v}$ and up $\mathbf{u}$ vectors:
\begin{align} \label{eq:b:lookrot}
 LookRotation \left( \mathbf{v}, \mathbf{u} \right) &= Rot(\mathbf{R}(\mathbf{\Hat{t}}, \mathbf{\Hat{v}} \times \mathbf{\Hat{t}}, \mathbf{\Hat{v}}))
\end{align}
where:
\begin{alignat*}{3}
 \mathbf{\Hat{v}} = \frac{\mathbf{v}}{\|\mathbf{v}\|} & \qquad
 \mathbf{\Hat{u}} = \frac{\mathbf{u}}{\|\mathbf{u}\|} & \qquad
 \mathbf{\Hat{t}} = \frac{\mathbf{\Hat{v}} \times \mathbf{\Hat{u}}}{\|\mathbf{\Hat{v}} \times \mathbf{\Hat{u}}\|} &
\end{alignat*}
We sometimes use axis-angle rotation vectors (with the angle encoded as the length of the axis vector) to represent angular velocities. Quaternions can be converted into axis-angle representations as follows:
\begin{align} \label{eq:b:axisangle}
 AxisAngle (w, x, y, z)  = 2 \arccos{(w)} \frac{(x, y, z)}{\| (x, y, z) \|}
\end{align}

\subsection{Skeletal-based animation}

\begin{figure}[h]
  \centering
  \includegraphics[width=0.7\linewidth]{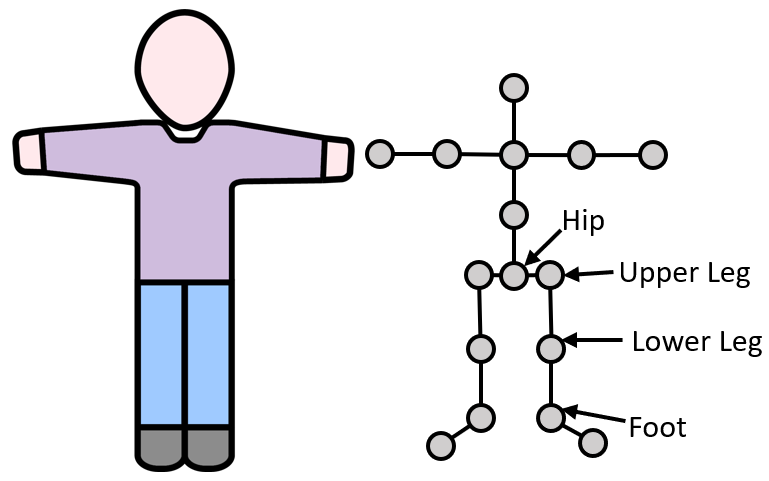}
  \caption[Virtual character skeleton]{\label{fig:back:skeleton} Representation of a humanoid skeleton used for animating virtual characters. Spheres represent joints.}
\end{figure}

Virtual characters are commonly rendered using triangular meshes, however, defining animations as the movement of triangles is not practical. Instead, we use a skeletal representation (see Figure~\ref{fig:back:skeleton}) composed of joints. Analogous to the human body, joints represent the articulated parts of the body. Mesh vertices are associated with one or more joints, when a joint moves, all vertices associated move as well.

Each joint $j$ is defined in the local frame of its parent, thus, it stores its local offset $\mathbf{t}_j \in \mathbb{R}^3$ and local rotation $\mathbf{q}_j \in \mathbb{R}^4$ represented as a quaternion. We can obtain the transformation matrix $\mathbf{P}_j \in \mathbb{R}^{4x4}$ that transforms from the local space defined by joint $j$ to the local space defined by its parent $p(j)$:
\begin{align}
    \mathbf{P_j} = 
    \begin{pmatrix}
        \mathbf{R}(\mathbf{q}_j) & \mathbf{t}_j \\
        0 & 1 \\
    \end{pmatrix}
\end{align}
where $\mathbf{R}(\mathbf{q}_j)$ is the rotation matrix obtained from the quaternion representation as Eq.~\ref{eq:b:quat_to_mat}. 

Obtaining the world position of a joint is commonly known as Forward Kinematics (FK) and consists of multiplying the transformation matrices of all parents and the joint itself. For instance, as shown in Figure~\ref{fig:back:skeleton}, to obtain the foot position, we multiply the Hip joint transformation matrix by the Upper Leg, then Lower Leg and finally the Foot. Formally:
\begin{align}
    FK(j) = \mathbf{P}_{root} \dots \mathbf{P}_{p(p(j))} \mathbf{P}_{p(j)} \mathbf{P}_j
\end{align}
\chapter{Animation System Implementation} \label{chap:mm}

In this chapter, we describe the animation system we have developed based on Motion Matching and all methods involved in synthesizing the final animation. Finally, we describe critical implementation details.

\begin{figure}[h]
  \centering
  \includegraphics[width=1.0\linewidth]{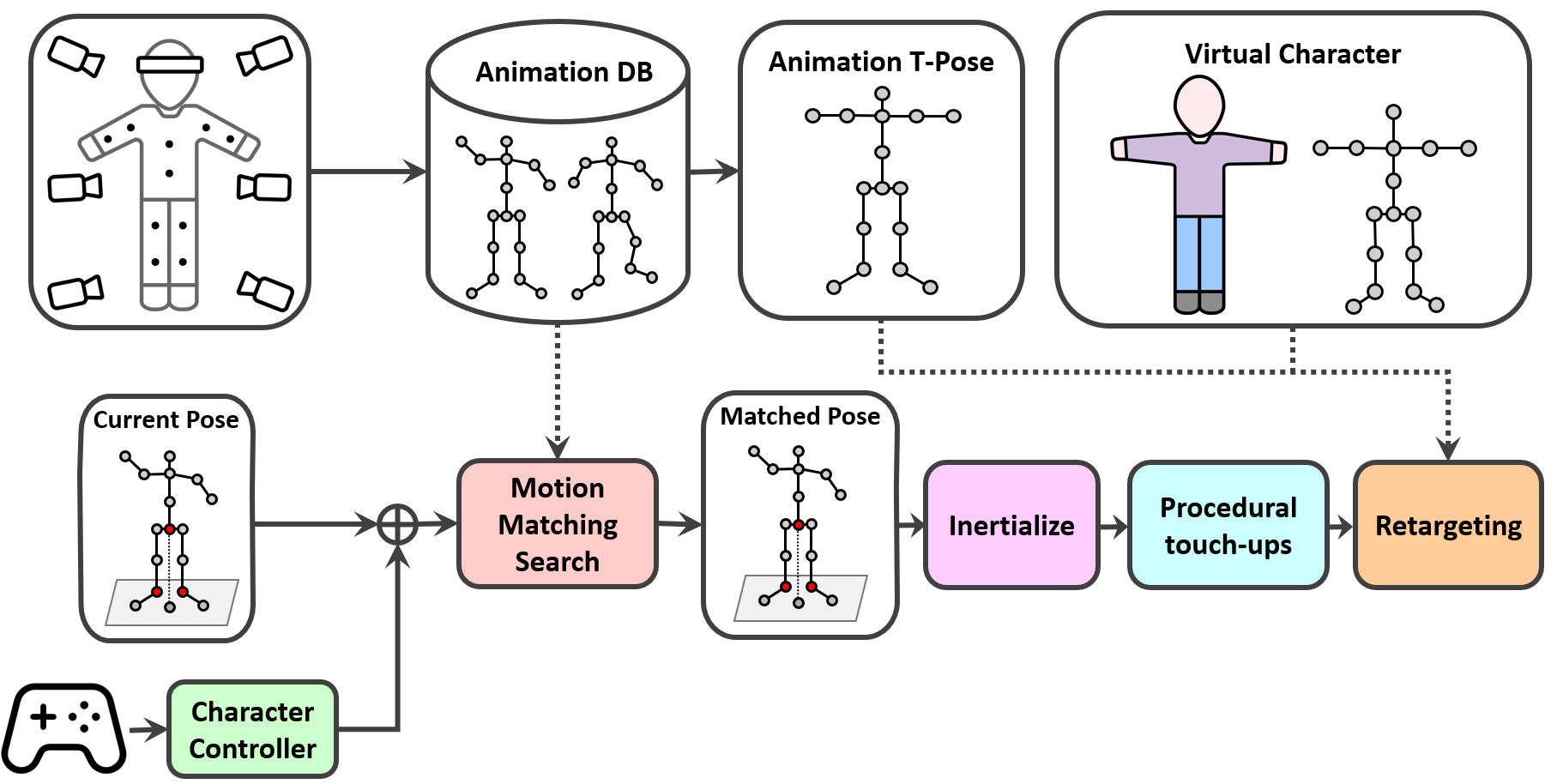}
  \caption[Animation System Pipeline]{\label{fig:mm:anim_pipeline} Animation pipeline for Motion Matching. First, the current pose and the target motion are extracted and used to create a feature vector for the Motion Matching Search. The returned pose is the best candidate that follows the user's input and minimizes the change in pose. The pose is blended using Inertialization to avoid sudden changes. Next, retargeting is applied to animate the virtual character, which may have a different skeleton. Finally, some procedural touch-ups enhance the final result. }
\end{figure}

Motion Matching is a data-driven algorithm to animate virtual characters with minimal manual setup and seamless integration of different types of movements. Unlike traditional methods, Motion Matching does not require manually setting up a state machine with all kinds of movements and defining how to transition between them. Instead, it works by searching over an animation database for the best pose sequence that matches the current pose and the intended trajectory of the character. It was initially presented by \citet{buttner2015} as a greedy approximation to the Motion Fields method presented by \citet{lee2010}, and it was further developed for Ubisoft's game \emph{For Honor} \citep{clavet2016}. Recently, \citet{holden2020} has provided a state-of-the-art implementation used in AAA game productions which we use as the primary reference for our implementation. An overview of all systems involved in the animation system can be seen in Figure~\ref{fig:mm:anim_pipeline}.

Animation data is a critical aspect of Motion Matching. Animations are usually recorded using motion capture to obtain high-fidelity motion; however, it is also possible to use hand-made animations in any animation software such as Blender or Autodesk Maya. Motion Matching does not synthesize new poses, instead, at runtime, it searches for the best pose to apply given the current pose and user input. Poses are played in the original order as captured, however, every a few frames, a search is performed to obtain a new sequence of poses that best matches the predicted user trajectory. Since there may be noticeable changes in the virtual character's pose, it is blended using Inertialization \citep{bollo2017}. Then, the pose is retargeted into the virtual character's skeleton, and some procedural touch-ups help polish the final animation.

Although Motion Matching is essentially a search in an ordered array and does not use complex abstractions, it provides numerous advantages compared to previous approaches based on Motion Graphs: no graph construction allows for more flexibility and quicker response to user input; it is compatible with existing systems such as physically-based characters since it can handle arbitrary poses; data is represented in arrays of floats contiguous in memory, therefore standard compression and optimization techniques can be used. 

\section{Motion Matching} \label{sec:mm:mm}

The core of a character animation system can be seen as a process that outputs a sequence of poses based on some input. Two main components are involved in this process: the character controller (commonly known as the gameplay code) and the animation engine. The former is responsible for receiving the user's input and sending to the animation engine all information needed to animate the virtual character: speed, movement direction, gaits, gender, and look at direction, among others. User input may not be necessary in some cases, such as when moving a virtual agent. The animation engine receives all information and outputs a pose for every frame. Both components are separated but must be communicated and designed to work together. In this section, we focus on the animation engine while the character controller is explained in Section~\ref{sec:mm:character_controller}.

Traditionally, a state machine controls the animation engine, as shown in Figure~\ref{fig:intr:animation_controller}. A node can be seen as an animation clip, e.g., idle, while edges contain the conditions that need to be satisfied to change the node, e.g., after a certain amount of time or when a Boolean is set to true from the character controller. Nodes usually contain nested state machines or blending spaces to support complex animations. Moreover, poses can be combined by running several state machine layers. For instance, one layer may control the lower body while another layer the upper body. There are several methods for combining layers, such as override, addition, and difference. Altogether, state machines are challenging to create and maintain. For instance, if we want to add an injured state, the number of nodes to incorporate is enormous; the character should be able to transition from almost any state to injured. 

\begin{figure}[h]
  \centering
  \includegraphics[width=1.0\linewidth]{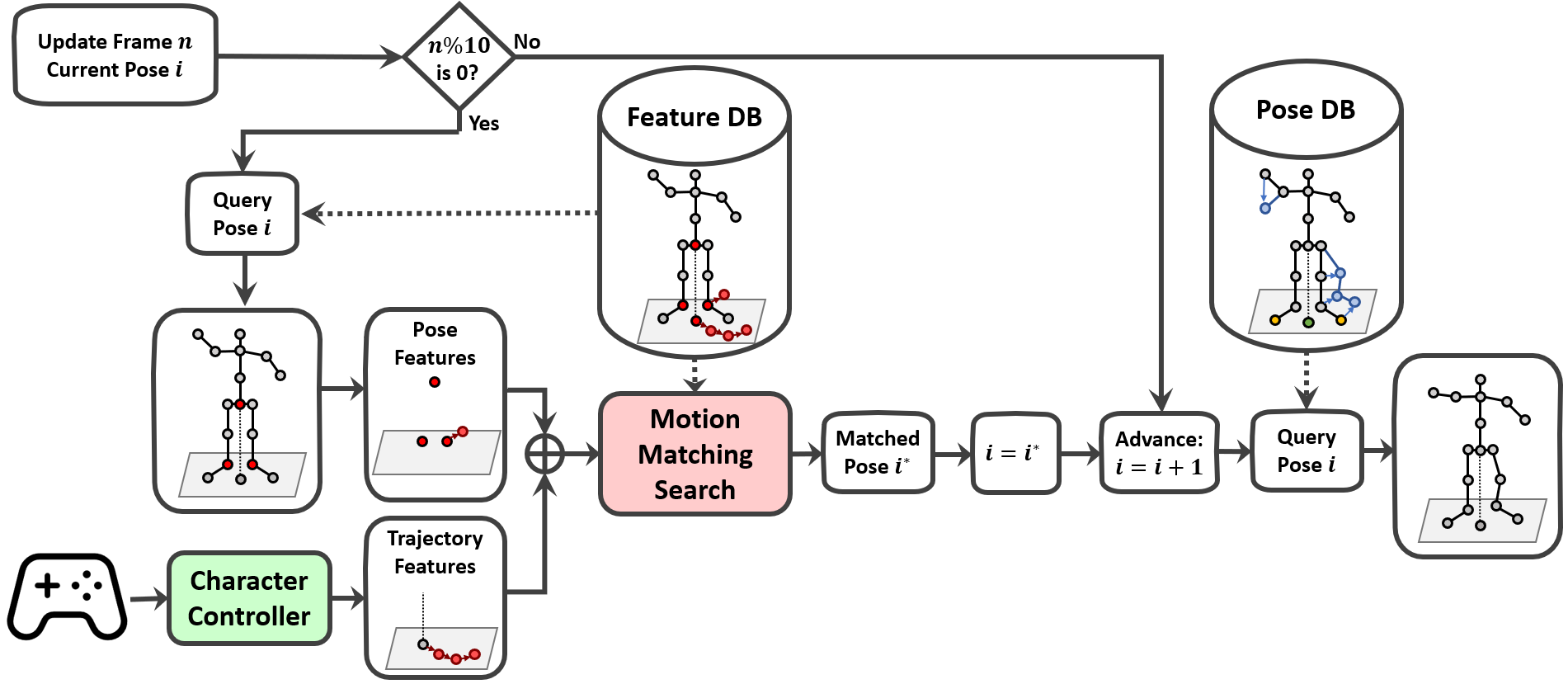}
  \caption[Motion Matching Search]{\label{fig:mm:search} Poses are played consecutively each frame. Every a few frames (e.g., 10) a search is performed to find the sequence of poses that best matches the predicted user trajectory and the current pose.}
\end{figure}

Motion Matching tackles this issue by formulating the problem of obtaining a pose as a search over an animation database. The gameplay code keeps all logic and state and sends similar information as before. The animation database can be tagged, thus, relevant movements can be quickly recovered depending on the gait, gender, or other discrete properties. Continuous information such as the trajectory or directions is combined to construct a feature vector that, together with features from the current pose, is used to search the animation database for the best match. An overview of this process is shown in Figure~\ref{fig:mm:search}.

\subsection{Pose database} \label{sec:mm:pose_database}

\begin{figure}[h]
  \centering
  \includegraphics[width=1.0\linewidth]{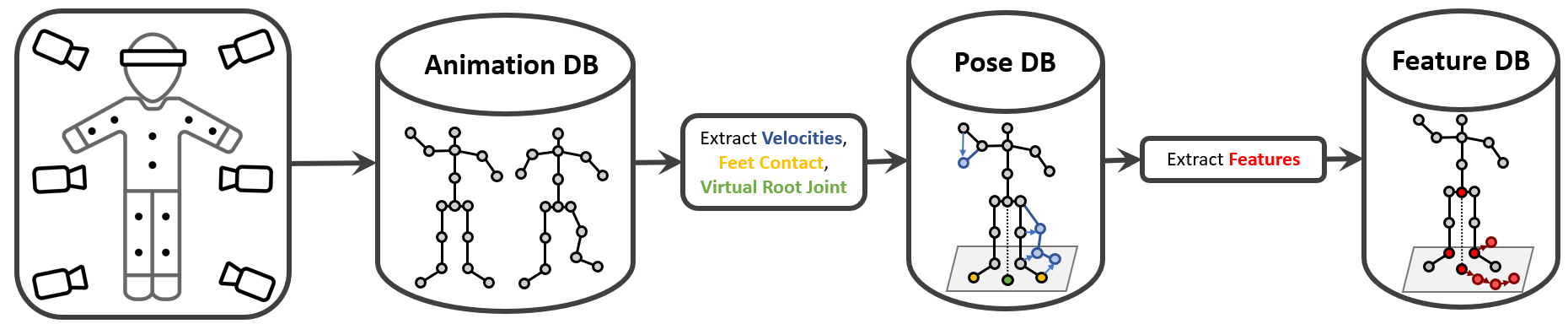}
  \caption[Animation, Pose and Feature databases]{\label{fig:mm:databases} All poses are contained in the animation database which is usually captured using motion capture systems. The pose database contains poses plus some additional information. Finally, the feature database only contains the required features for the Motion Matching search.}
\end{figure}

Motion Matching searches over an animation database for the best match for the current avatar pose and the predicted trajectory. Since Motion Matching does not create new poses, the animation database is an essential component that determines the quality of the final animations and can include one or more motion capture files. We used the Xsens motion capture system to record the animation database, exported as BVH's files (see Section~\ref{sec:mm:bvh}). These files were pre-processed to create the pose database, which is a vector containing all poses in the same order as they appear in the motion capture files but with additional information. This process is shown in Figure~\ref{fig:mm:databases}. Each pose $\mathbf{y}$ is defined as follows:
\begin{equation} \label{eq:mm:pose_vector}
    \mathbf{y} = \left( \mathbf{y^p}, \mathbf{y^r}, \mathbf{y^v}, \mathbf{y^w}, \mathbf{y^c} \right)
\end{equation}
where $\mathbf{y^p} = (\mathbf{y^p_0}, \dots, \mathbf{y^p_J})$ are the local joint positions and $J$ the number of joints, $\mathbf{y^r} = (\mathbf{y^r_0}, \dots, \mathbf{y^r_J})$ are the local joint rotations, $\mathbf{y^v} = (\mathbf{y^v_0}, \dots, \mathbf{y^v_J})$ are the local joint velocities, $\mathbf{y^w} = (\mathbf{y^w_0}, \dots, \mathbf{y^w_J})$ are the local joint angular velocities and $\mathbf{y^c} = (y^c_0, y^c_1)$ contains two Boolean values indicating whether the left and right foot are in contact with the ground.

Here, we describe how to construct the pose vector $\mathbf{y}$. The avatar returned by Xsens has the hip joint as the root of the skeleton. Therefore, we add a virtual root joint to the skeleton that indicates the position $\mathbf{p}$ and orientation $\mathbf{r}$ of the character. The virtual root joint will be necessary to define the character space later. It is created by projecting the hip joint onto the floor plane, and its coordinate frame is defined by the projected hip forward direction, the vertical world vector and their cross product. Then the hip joint is transformed into the virtual root space. Formally, we construct $\mathbf{y^p}$ and $\mathbf{y^r}$ from the animation database local joint positions $\mathbf{a^p} = (\mathbf{a^p_0}, \dots, \mathbf{a^p_{J-1}})$ and local joint rotations $\mathbf{a^r} = (\mathbf{a^r_0}, \dots, \mathbf{a^r_{J-1}})$:
\begin{alignat}{3}
    \mathbf{y^p} &= (\mathbf{p}, \; & \mathbf{(a^p_0)'}, \; & \mathbf{a^p_1}, \dots, \mathbf{a^p_{J-1}}) \\
    \mathbf{y^r} &= (\mathbf{r}, \; & \mathbf{(a^r_0)'}, \; & \mathbf{a^r_1}, \dots, \mathbf{a^r_{J-1}})
\end{alignat}
where (rotations are represented as quaternions, $LookRotation$ is defined as in Eq.~\ref{eq:b:lookrot}, $Rot$ as in Eq.~\ref{eq:b:rot}, $\Vec{up} = (0,1,0)$ and $\Vec{fw} = (0,0,1)$):
\begin{align}
 \mathbf{p}  &=  \mathbf{a^p_0} - proj_{\Vec{up}}(\mathbf{a^p_0}) \\
 \mathbf{r}  &=  LookRotation \left( \mathbf{f} - proj_{\Vec{up}}(\mathbf{f}), \Vec{up} \right) \\
 \mathbf{f}  &=  Rot_{\mathbf{a^r_0}}(\Vec{fw}) \\
 \mathbf{(a^p_0)'} &=  Rot_{\mathbf{r}^{-1}}(\mathbf{a^p_0} - \mathbf{p}) \\
 \mathbf{(a^r_0)'} &=  \mathbf{r}^{-1} \mathbf{a^r_0}
\end{align}

Once local positions and rotations are computed, we calculate the local joint velocities $\mathbf{y^v}$ and angular velocities $\mathbf{y^w}$ via finite difference. Here, the super index $(i)$ refers to the $i$-th example from the database, and $\Delta t$ is the time between two poses:
\begin{align} \label{eq:mm:finite_difference}
    (\mathbf{y^v})^{(i)} = \frac{(\mathbf{y^p})^{(i)} - (\mathbf{y^p})^{(i-1)}}{\Delta t}
\end{align}
And for angular velocities the idea is the same, but quaternions represent rotations, and axis-angle rotation vectors (with the angle encoded as the length of the axis vector) represent angular velocities:
\begin{align}
    (\mathbf{y^w})^{(i)} = \frac{AxisAngle \left( (\mathbf{y^r})^{(i)} ((\mathbf{y^r})^{(i-1)})^{-1}  \right) }{\Delta t}
\end{align}
where $AxisAngle$ is defined as in Eq.~\ref{eq:b:axisangle}.

Finally, to compute $\mathbf{y^c}$, we check whether the velocity of the toes joint is close to zero. Then, the result is smoothed by a median filter with a window of radius 6. The filter acts as a low-pass filter and prevents constant changes in contact information.

\subsection{Feature database}
Notice that $\mathbf{y}$ contains data relative to a given animation frame. Since we wish to search for the best sequence of poses, we need to add temporal information. Therefore, instead of directly using the pose database when searching for a new sequence of poses, we compute a new database with the main features defining the motion. In this section, we describe the features needed for locomotion \citep{clavet2016}, which is the primary purpose of this work and Motion Matching systems; however, other features could be included in the vector and the formulation of the method will remain. Our implementation can also handle a large set of user-defined features.

\begin{figure}[h]
  \centering
  \includegraphics[width=1.0\linewidth]{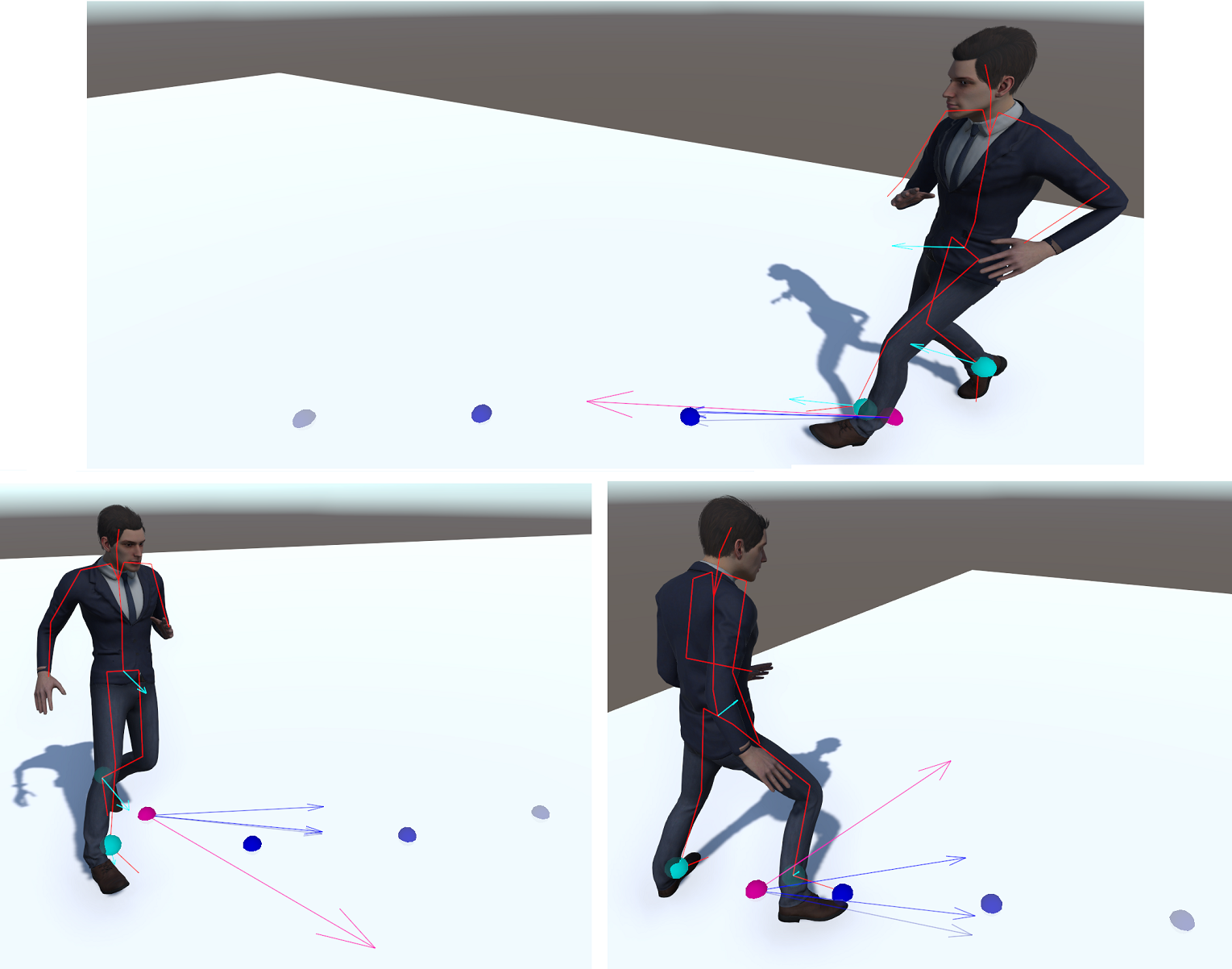}
  \caption[Pose and Trajectory Features]{\label{fig:mm:features} Pose and trajectory features for locomotion in different poses. Cyan spheres are the feet joints positions, cyan arrows are the feet and hip velocities. Blue spheres and arrows are the future positions and directions if poses are played consecutively. In magenta the current position and direction of the virtual root joint.}
\end{figure}

We compute a feature vector $\mathbf{z} \in \mathbb{R}^{27}$ for each pose $\mathbf{y}$. This feature vector combines two types of information: the current pose and the trajectory. When comparing feature vectors, the former ensures no significant changes in the pose and thus smooth transitions; the latter drives the animation towards our target trajectory. Feature vectors are defined as follows:
\begin{equation} \label{eq:mm:z}
    \mathbf{z} = \left( \mathbf{z^v}, \mathbf{z^l}, \mathbf{z^p}, \mathbf{z^d} \right)
\end{equation}
where $\mathbf{z^v}, \mathbf{z^l}$ are the current pose features and $\mathbf{z^p}, \mathbf{z^d}$ are the trajectory features. More precisely, $\mathbf{z^v} \in \mathbb{R}^{3\times3}$ are the velocities of the feet and hip joints, $\mathbf{z^l} \in \mathbb{R}^{3\times2}$ are the positions of the feet joints, $\mathbf{z^p} \in \mathbb{R}^{2\times3}$ and $\mathbf{z^d} \in \mathbb{R}^{2\times3}$ are the future 2D positions and 2D orientations of the character $0.33$, $0.66$ and $1.00$ seconds ahead. Figure~\ref{fig:mm:features} visually shows the different features for locomotion. Note that more than three predicted points could be used depending on the application requirements. For example, 6 trajectory positions could be used: three in the past $-1.00$, $-0.66$ and $-0.33$ seconds, and three predicted $0.33$, $0.66$ and $1.00$ seconds. Trajectory features are projected onto the ground and all features are local to the virtual root joint, i.e., in character space. Each feature is also normalized by subtracting its mean and dividing by the standard deviation.

The vector $\mathbf{z^l}$ is computed by applying forward kinematics (FK) using the local positions $\mathbf{y^p}$ and rotations $\mathbf{y^r}$ from the pose vector $\mathbf{y}$. When computing FK, if the position and rotation of the virtual root joint are not applied, the resulting position will already be in character space. The velocity $\mathbf{z^v}$ can be computed from the finite difference of $\mathbf{z^l}$, similarly as in Eq.~\ref{eq:mm:finite_difference}.

\newpage
Trajectory features $\mathbf{z^p}, \mathbf{z^d}$ are computed by sampling the position and direction of the virtual root joint in consecutive poses and making them local to the current character local frame. For instance, if we assume the application is running at 60 frames per second:
\begin{align}
    (\mathbf{z^p})^{(i)} &= \left( Rot_{\mathbf{r'}}(\mathbf{p}^{(i+20)} - \mathbf{p}^{(i)}), \; Rot_{\mathbf{r'}}(\mathbf{p}^{(i+40)} - \mathbf{p}^{(i)}), \; Rot_{\mathbf{r'}}(\mathbf{p}^{(i+60)} - \mathbf{p}^{(i)}) \right) \\
    (\mathbf{z^d})^{(i)} &= \left( Rot_{\mathbf{r'}\mathbf{r}^{(i+20)}}(\Vec{fw}), \; Rot_{\mathbf{r'}\mathbf{r}^{(i+40)}}(\Vec{fw}), \; Rot_{\mathbf{r'}\mathbf{r}^{(i+60)}}(\Vec{fw}) \right)
\end{align}
% local explain here TODO
where $\mathbf{p} = \mathbf{y^p_0}$, $\mathbf{r} = \mathbf{y^r_0}$ and  $\mathbf{r'} = \mathbf{r}^{-1}$. Note that superindex $(i)$ refers to the $i$-th example from the database, and assigning a 3D vector to a 2D vector removes the vertical component. 

\subsection{Search} \label{sec:mm:search}
At runtime, we perform a Motion Matching search every a few frames (e.g., 10 frames). At a given update, the character is at a certain pose described by the feature vector $\mathbf{z}$, and we want to search for the sequence of poses that best matches the predicted user's trajectory. To do so, we create a query feature vector $\mathbf{\Hat{z}}$ defined as in Eq.~\ref{eq:mm:z} and update it before every search. The pose features $\mathbf{\Hat{z}^v}$ and $\mathbf{\Hat{z}^l}$ are simply copied from the current feature vector $\mathbf{z}$. Trajectory features $\mathbf{\Hat{z}^p}$ and $\mathbf{\Hat{z}^d}$ are computed by the character controller (see Section~\ref{sec:mm:trajectory_prediction}).

The query vector $\mathbf{\Hat{z}}$ is used to search in the feature database for the closest vector:
\begin{align}
    i^* = \argmin_{i} \| \mathbf{\Hat{z}} - \mathbf{z}^{(i)} \|^2
\end{align}
Since features are normalized, we can add weights $(\beta^v, \beta^l, \beta^p, \beta^d)$ to modify the priority of some features:
\begin{equation}
    i^* = \argmin_{i} \left( \begin{array}{l@{}ll}
                        \beta^v & \|\mathbf{\Hat{z}^v} - (\mathbf{z^v})^{(i)}\|^2 & + \\
                        \beta^l & \|\mathbf{\Hat{z}^l} - (\mathbf{z^l})^{(i)}\|^2 & + \\
                        \beta^p & \|\mathbf{\Hat{z}^p} - (\mathbf{z^p})^{(i)}\|^2 & + \\
                        \beta^d & \|\mathbf{\Hat{z}^d} - (\mathbf{z^d})^{(i)}\|^2 & 
                        \end{array}\right)
\end{equation}
As modifying each weight individually may be counterintuitive, we also define two parameters: $\beta^r$ multiplies the trajectory features, thus controlling the responsiveness; $\beta^q$ controls the quality of the animation by multiplying the pose features:
\begin{equation} \label{eq:mm:search_weights}
    i^* = \argmin_{i} \left( \begin{array}{l@{}ll}
                        \beta^q \beta^v & \|\mathbf{\Hat{z}^v} - (\mathbf{z^v})^{(i)}\|^2 & + \\
                        \beta^q \beta^l & \|\mathbf{\Hat{z}^l} - (\mathbf{z^l})^{(i)}\|^2 & + \\
                        \beta^r \beta^p & \|\mathbf{\Hat{z}^p} - (\mathbf{z^p})^{(i)}\|^2 & + \\
                        \beta^r \beta^d & \|\mathbf{\Hat{z}^d} - (\mathbf{z^d})^{(i)}\|^2 & 
                        \end{array}\right)
\end{equation}
The returned pose is not necessarily the continuation of the current pose in the animation database. Although we added temporal information to the feature vectors to prime smooth transitions, there may still be some noticeable changes between poses, especially for motions not included in the database. Therefore, we apply Inertialization blending \citep{bollo2017} to smooth the transitions (see Section~\ref{sec:mm:inertialization}).

Since poses are ordered in the pose database, we store an index indicating the pose for the current frame. This index is incremented every frame to create the animation. The search is performed at the beginning of the frame, and the pose features for the query vector are extracted from the current pose (which is still the same as the previous frame), thus, we still increment the index. In other words, ideally, the search returns another pose with the same pose features as the current one but with trajectory features following the predicted user's trajectory. Then, we must always increment the index to advance in the animation. An overview of the search is shown in Figure~\ref{fig:mm:search}.

\section{Animation database}
This section describes and discusses the steps needed to capture and use the animation database. Correctly creating the animation database is essential since the quality of the data will drastically change the resulting animation. First, one or several motion capture files are used to form the animation database; this data is processed and stored in the Biovision hierarchical data (BVH) file format. We implemented an importer of BVH in Unity and used it to extract all poses to create the pose database. The final pose is retargeted into the virtual character's skeleton. In the following sections, we explain these processes in detail.

\subsection{Motion capture} \label{sec:mm:mocap}
\begin{figure}[h]
  \centering
  \includegraphics[width=1.0\linewidth]{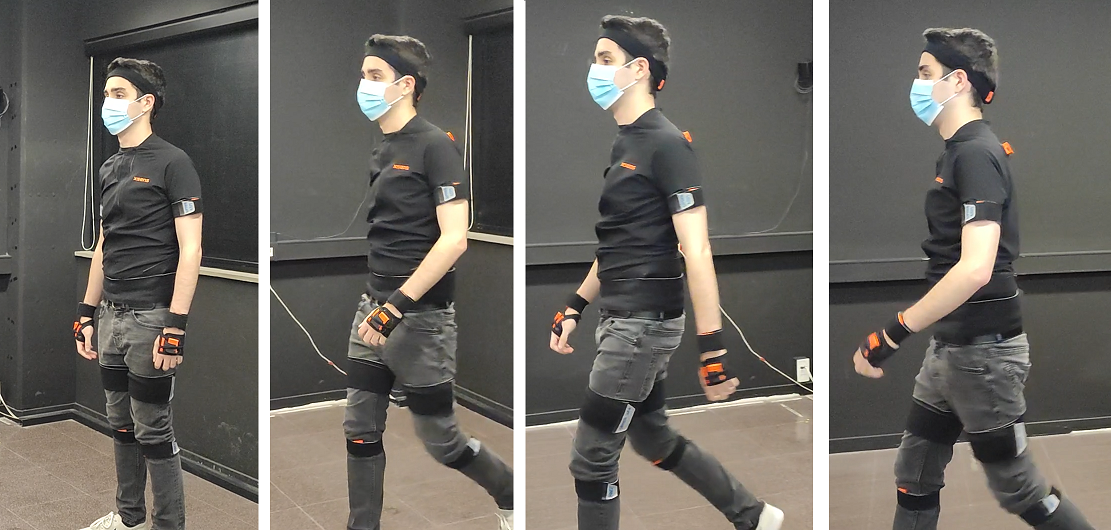}
  \caption[Motion Capture]{\label{fig:mm:mocap} A user with Xsens Awinda motion capture suit recording a locomotion sequence.}
\end{figure}

Motion capture (sometimes referred to as mocap) is the set of techniques to record objects or people's movements. It is widely used in entertainment, sports, medical applications, and the validation of robots, among others. Primarily, mocap systems can be divided into two categories:
\begin{itemize}
	\item \textbf{Optical systems} use images captured from sensors or cameras to triangulate the position of multiple sensors or markers attached to an actor. These systems are typically expensive due to the need for numerous sensors or cameras to avoid occlusions and cover the tracking space. Recently, deep learning-based methods are able to recover 3D poses directly from one or more cameras without the need for specific markers or suits; however, the final quality is still not as good as traditional systems.
	\item \textbf{Non-optical systems} do not use external cameras or sensors to capture the actor's movement. Inertial systems are gaining popularity recently due to their versatility and easy setup. These systems use inertial measurement units (IMUs) that consist of a gyroscope, magnetometer and accelerometer to measure rotation. The software translates the rotation into a skeleton. The main drawback is lower positional accuracy and positional drift over time.
\end{itemize}
We use the Xsens Awinda system (see Figure~\ref{fig:mm:mocap}), which uses IMUs to record the actor's pose. We chose the Xsens system because it provides high-quality pose recording, an automatic post-processing system that ensures smooth and continuous motion and does not have occlusion problems. Moreover, the system can be easily set up (from 5 to 10 minutes), making it suitable for quick prototyping and research. 

\begin{figure}[h]
  \centering
  \includegraphics[width=1.0\linewidth]{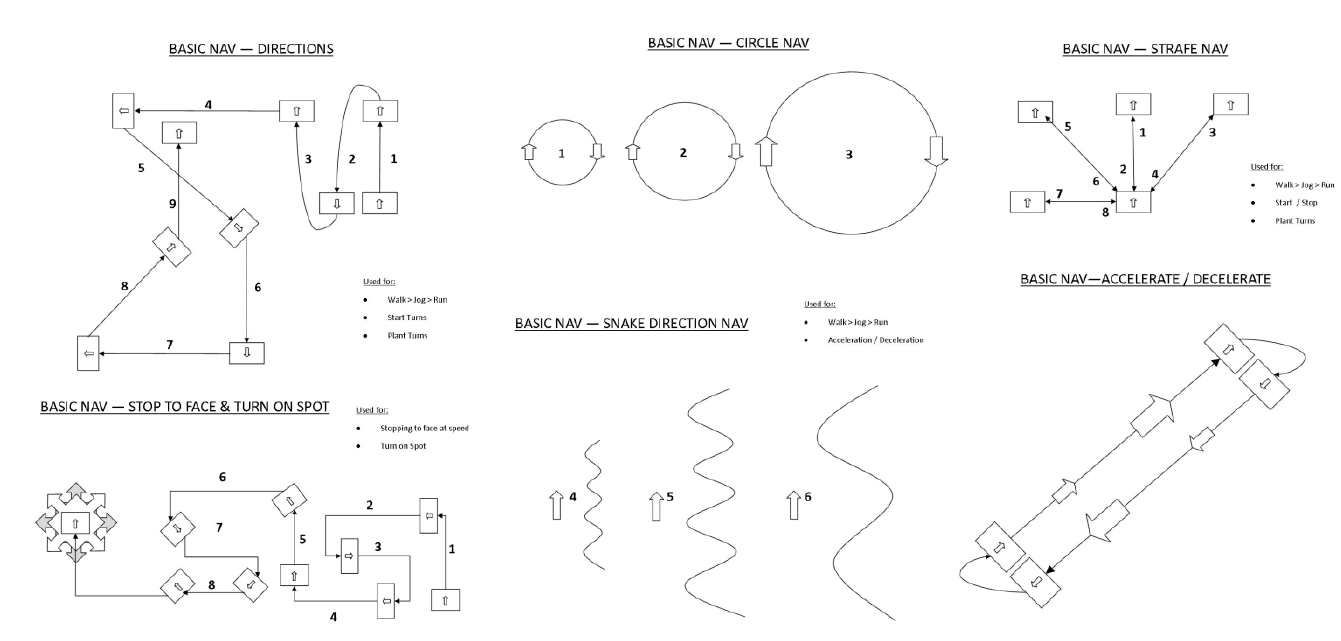}
  \caption[Dance Cards]{\label{fig:mm:dance_cards} Dance cards used to specify the motions to record during a mocap session for locomotion. They help obtain clean and appropriate data for Motion Matching. Source: \citep{kristjanzadziuk2016}.}
\end{figure}

\newpage
Motion Matching works by searching the animation database for the sequence that best matches the current pose and the predicted user trajectory. Therefore, it works with any database regarding its duration or type of movements. Nonetheless, following some simple rules will substantially improve the final quality of the animation:
\begin{itemize}
	\item Avoid undesired and ambiguous movements. One advantage of Motion Matching is that it retains high-quality mocap data, including small high-frequency signals. For instance, if the actor slightly turns their heads while walking, this will be represented in the final animation. The same situation occurs when repeating a movement, e.g., walking forward with different arms waving. Motion Matching will interpret both actions as walking forward, but their arms waving will differ; when animating the virtual character, the system may mix both animations instead of following one of the sequences.
	\item Variety of transitions. The animation database should cover a large spectrum of transitions, for instance, suppose it only represents right and left turns of 90 degrees, if the virtual character turns 45 degrees to the right, Motion Matching will not find a suitable sequence and may have to perform 90 degrees turn or stop in the middle with an abrupt transition.
	\item Trajectories are usually evaluated 1 second in the future. If the actor wants to record a walking forward action followed by an idle state, they should remain at least 1 second idle after walking to facilitate the search at runtime.
\end{itemize}
In order to provide an easy procedure for capturing movements while following the mentioned rules, \citet{kristjanzadziuk2016} presented the concept of Dance Cards (see Figure~\ref{fig:mm:dance_cards}). The idea is to specify the type of motions to capture before the mocap session. Dance cards make the process reproducible and can be improved iteratively.

\subsection{Biovision hierarchical data (BVH)} \label{sec:mm:bvh}
% explain format... conversions between euler, quaternion... etc.

\begin{figure}[h]
  \centering
  \includegraphics[width=1.0\linewidth]{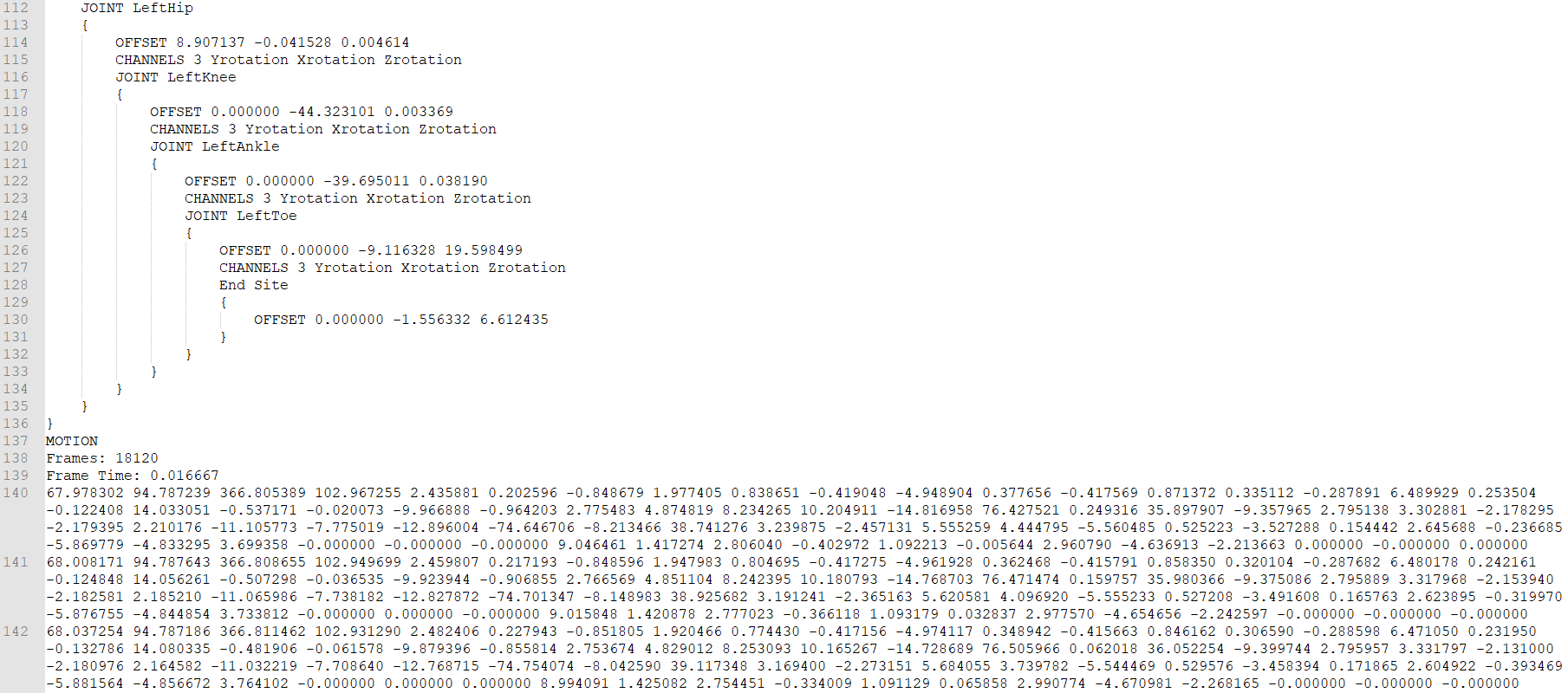}
  \caption[Biovision hierarchical data (BVH)]{\label{fig:mm:bvh} Biovision hierarchical data (BVH) mocap file data. The skeleton is specified (top) followed by the motion recorded frame by frame (bottom). }
\end{figure}

Once the Xsens animation software captures and processes the motion, it stores the information in a BVH file (see Figure~\ref{fig:mm:bvh}), commonly used for mocap as poses and motion are stored compactly.

The file starts by hierarchically defining the skeleton with the keyword \emph{HIERARCHY} followed by \emph{ROOT} and the name of the root joint. Then, brackets are used to define the root's children's joints. Each joint is similarly defined but with the keyword \emph{JOINT}. For each joint (or the root), the local offset with respect to its parent is specified with the keyword \emph{OFFSET} followed by three scalar values representing the 3D local position. Our system assumes that offsets are fixed, so the distances are preserved. Thus, each joint contains the keyword \emph{CHANNELS 3 Yrotation Xrotation Zrotation} specifying that each frame will store three values to represent the joint's local rotation. Other ordering may be set (e.g., \emph{CHANNELS 3 Xrotation Yrotation Zrotation}). For the root, we also store the position of the character in world space (commonly known as root motion) \emph{CHANNELS 6 Xposition Yposition Zposition Yrotation Xrotation Zrotation}. The keyword \emph{End Site} is used when a joint has no children.

After the skeleton definition, the keyword \emph{MOTION} indicates the beginning of the pose data. First, \emph{Frames:} and \emph{Frame Time:} store the number of frames and the time in seconds per frame, respectively. The rest of the file contains the actual motion data. Each line is one sample, and the numbers appear in order of the channel specification, reading the hierarchy in preorder.

We developed a custom BVH importer as Unity does not provide a built-in method. Apart from parsing the file to store the skeleton and motion data, an essential step is transforming the data from BVH's right-handed to Unity's left-handed representation. We also transform the rotations representation from Euler angles to quaternions to facilitate the pose and feature databases creation.

Firstly, the BVH's root world position $\mathbf{p} = (p_x, p_y, p_z)$ is easily transformed into a left-handed representation $\mathbf{p'} = (p'_x, p'_y, p'_z)$ as follows:
\begin{align}
\begin{split}
    p'_x &= p_x \\
    p'_y &= p_y \\
    p'_z &= -p_z
\end{split}
\end{align}
considering BVH's positive Z axis is Unity's negative Z axis and other axes remain the same.

\newpage
Secondly, local rotations are specified in the BVH file as Euler angles in an arbitrary order. The order is essential, otherwise, when constructing the quaternion, we may obtain different orientations. We will assume the rotation order is Y, X and Z as returned by Xsens. A rotation is represented in a quaternion form by encoding axis-angle information as follows: 
\begin{align}
    \mathbf{q} = QuatRot(\theta, (x, y, z)) = (\cos{(\theta / 2)},\, \sin{(\theta / 2)} (x, y, z))
\end{align}
where $(x, y, z)$ is the unit axis of rotation rotated by the angle $\theta$. Therefore, given the three scalars $(r_y, r_x, r_z)$ from the BVH file, the transformed quaternion representation $\mathbf{q'}$ is defined:
\begin{align} \label{eq:mm:euler_to_quaternion}
    \mathbf{q'} = QuatRot(-r_y, (0, 1, 0)) QuatRot(-r_x, (1, 0, 0)) QuatRot(r_z, (0, 0, 1))
\end{align}
where rotations will be applied in order from left to right. Y and X-axis angles are negated (following the right and left-hand rules) to transform from BVH's right-handed into Unity's left-handed representation. BVH's with different rotation ordering can be transformed by changing the order of the multiplications in Eq.~\ref{eq:mm:euler_to_quaternion}. 

\subsection{Skeleton retargeting} \label{sec:mm:retargeting}
Our animation system selects a pose from the animation database for every frame and uses it to rotate the joints of a virtual character's skeleton. However, we cannot apply the pose to an arbitrary skeleton, even if they have the same number of joints, their initial position with respect to their parents is essential. Consider two skeletons, their joints are positioned such that if we apply the identity rotation to all joints, one skeleton has the head rotated 90 degrees to the right and the other looks at the front. If we want to represent the pose where the head is rotated 90 degrees, the former will have the identity rotation for the head joint, while the latter will have a 90 degrees rotation. In other words, the same pose is achieved with different rotations depending on the initial position of the joints.

\begin{figure}[h]
  \centering
  \includegraphics[width=1.0\linewidth]{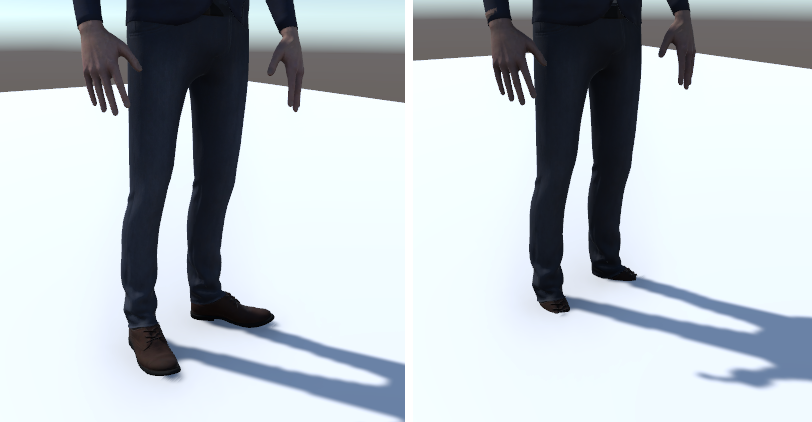}
  \caption[Skeleton Differences]{\label{fig:mm:skeleton_diff} Left: original skeleton used when capturing the animation database. Right: skeleton with longer legs. Although retargeting will make both skeletons have the same pose, it does not consider differences in joint lengths and thus the feet appear under the floor. }
\end{figure}

The animation will work when applied to the same skeleton in the mocap data. However, we want to use our system for different virtual characters. In that case, we need to perform retargeting by transforming the joint rotations so that the target skeleton has the desired pose. Note that our goal is applying poses to any skeleton regardless of its initial state; however, it is out of the scope of this work to retarget poses when distances between joints are different. For example, longer legs may cause the feet to intersect with the floor (see Figure~\ref{fig:mm:skeleton_diff}).

When the user imports a virtual character or an animation database, it has to specify the correspondence of each joint with a general humanoid skeleton convention. We use the already well-known convention by Unity\footnote{\href{https://docs.unity3d.com/ScriptReference/HumanBodyBones.html}{https://docs.unity3d.com/ScriptReference/HumanBodyBones.html}}. However, we do not use anything else from Unity's built-in animation system (as it is not suitable for Motion Matching). With this correspondence, we can relate joints between different skeletons.

Next, we require the T-Pose for each skeleton which is easy to obtain since most virtual characters are in T-Pose when no pose is applied, and mocap systems usually export a T-Pose in the first frame of the animation. Then, we have the source (animation database) and target (virtual character) skeletons at runtime. Given an arbitrary pose and joint, let $\mathbf{q^s}$ be the quaternion representing the rotation in world space for the source joint. We can compute the retargeted rotation $\mathbf{q^t}$ for the target joint with the following equation:
\begin{align}
    \mathbf{q^t} = \mathbf{q^s} \mathbf{q^{st}} \label{eq:mm:retargeting_1}
\end{align}
where $\mathbf{q^{st}}$ is the quaternion that rotates the local frame defined by $\mathbf{q^s}$ into the one defined by $\mathbf{q^t}$. Although $\mathbf{q^{st}}$ is unknown, we can compute it using the initial rotations when the skeletons are in T-Pose, which we represent with a 0 subindex:
\begin{align}
    \mathbf{q^t_0} &= \mathbf{q^s_0} \mathbf{q^{st}} \\ 
    \mathbf{q^{st}} &= (\mathbf{q^s_0})^{-1} \mathbf{q^t_0} \label{eq:mm:retargeting_2}
\end{align}
$\mathbf{q^{st}}$ can be obtained from Eq.~\ref{eq:mm:retargeting_2} and substituted into Eq.~\ref{eq:mm:retargeting_1} to obtain the final retargeted rotation:
\begin{align}
    \mathbf{q^t} &= \mathbf{q^s} (\mathbf{q^s_0})^{-1} \mathbf{q^t_0} \label{eq:mm:retargeting}
\end{align}
Thus, applying Eq.~\ref{eq:mm:retargeting} for each joint at an arbitrary pose will return the retargeted rotation for the virtual character.

\section{Character controller} \label{sec:mm:character_controller}
In Section~\ref{sec:mm:mm} we presented the animation engine based on Motion Matching, which is the system responsible for generating animations for virtual characters. However, another system needs to maintain the logic and state of the virtual character and communicate with the user's input or other systems (e.g., AI navigation). The character controller will be responsible for these tasks, as well as creating features that describe the desired motion.

Features sent by the character controller to the animation engine may include trajectories, gaits, and look at directions, among others. In this section, we focus on creating trajectories for locomotion (future positions and directions of the character), which is the primary purpose of this work. Other features such as gaits are typically implemented by tagging the animation database and restricting the search over the desired set of tags. 

To generate trajectories, we first need to specify the character's origin. A first approach could be to use the virtual root joint position from the virtual character and generate future positions and directions based on the user's input. However, the character's control will feel heavy and unpredictable because the virtual root joint depends exclusively on the motion capture data, which does not necessarily follow precisely the target future positions and directions.

Another approach is to maintain a separated position and orientation, named the simulated character, that will be code-driven by the character controller. The simulated character could then override the position and orientation of the virtual root joint. However, the problem is that all root motion information from the animation will be lost. Despite this, it is commonly used in many fast-paced games where responsiveness is more important than the animation quality. 

\begin{figure}[h]
  \centering
  \includegraphics[width=1.0\linewidth]{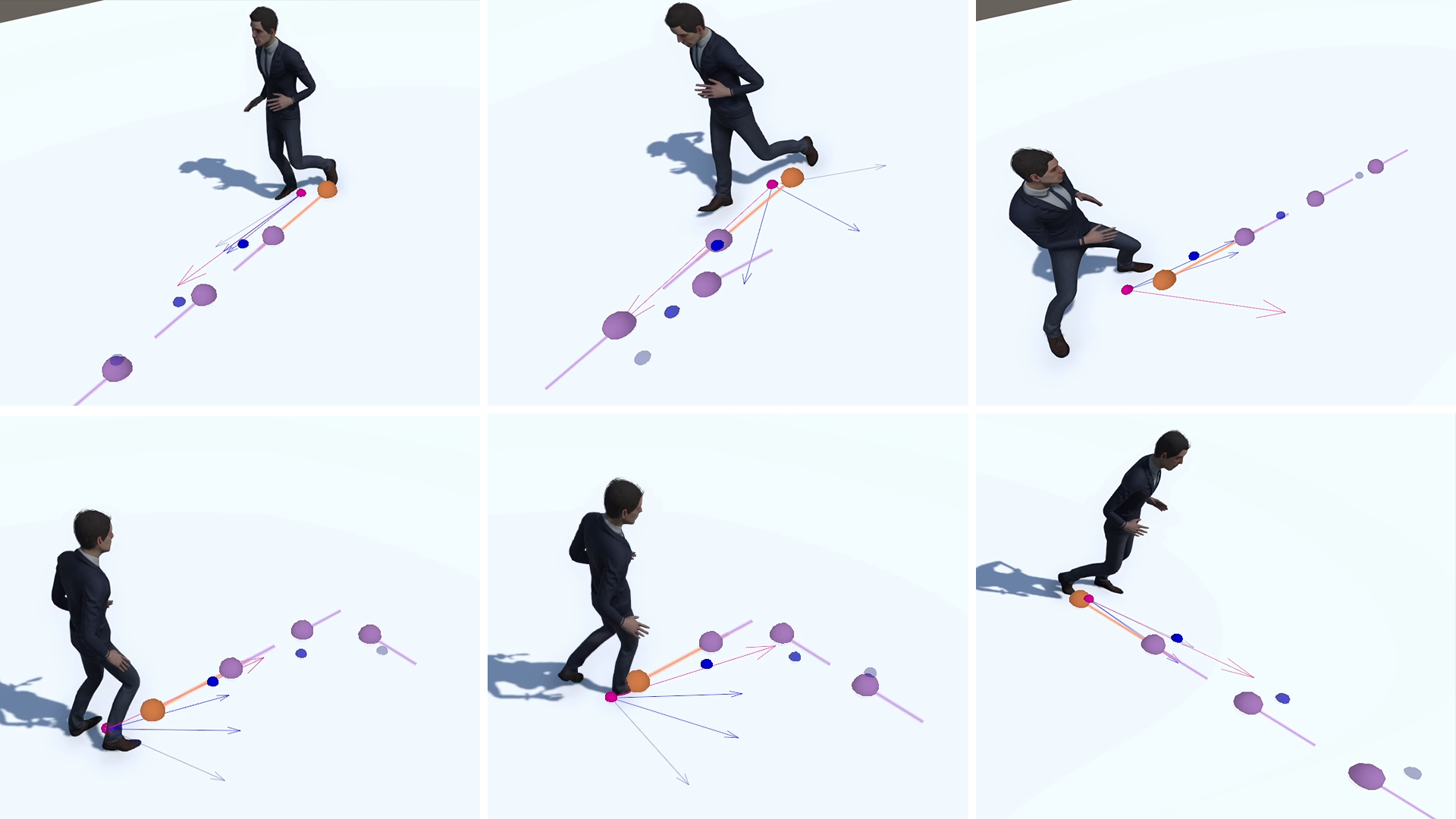}
  \caption[Trajectory Matching]{\label{fig:mm:path_mm} Magenta spheres and arrows are the virtual root joint positions and directions. Blue spheres and arrows are their future position and directions. The orange and purple spheres and lines are similarly defined for the code-driven simulated character. Notice how the blue trajectory features always try to match the purple simulated character trajectories. Sometimes, the virtual character deviates from the simulated character position because of the limited set of available animations.}
\end{figure}

Instead, as we are using Motion Matching, we can use an intermediate approach. On the one hand, the code-driven simulated character is moved by the character controller; on the other hand, the virtual root joint is animation-driven. Still, due to the trajectory features, it will always try to align with the simulated character. This approach retains responsive control while maintaining quality. Figure~\ref{fig:mm:path_mm} shows how Motion Matching always tries to align the virtual root joint (magenta and blue) with the simulated character (orange and purple).

\subsection{Trajectory prediction} \label{sec:mm:trajectory_prediction}
The main process of Motion Matching is the search over the feature database. It finds the closest feature vector to a query vector $\mathbf{\Hat{z}}$ defined as in Eq.~\ref{eq:mm:z}. The query vector has pose features $\mathbf{\Hat{z}^v}, \mathbf{\Hat{z}^l}$ that are automatically set from the current pose, and trajectory features $\mathbf{\Hat{z}^p}, \mathbf{\Hat{z}^d}$ that the character controller fills. This section describes how we create the trajectory features.

For locomotion, the trajectory is defined as the future positions and directions of the simulated character local to the virtual root joint. Trajectories should be created to resemble those found in the feature database, which is made with motion capture data. The most human-like these trajectories are, the best matching we will obtain when searching for sequences of poses. Thus, changes in positions and directions should be smooth and continuous. 

\begin{figure}[h]
  \centering
  \includegraphics[width=1.0\linewidth]{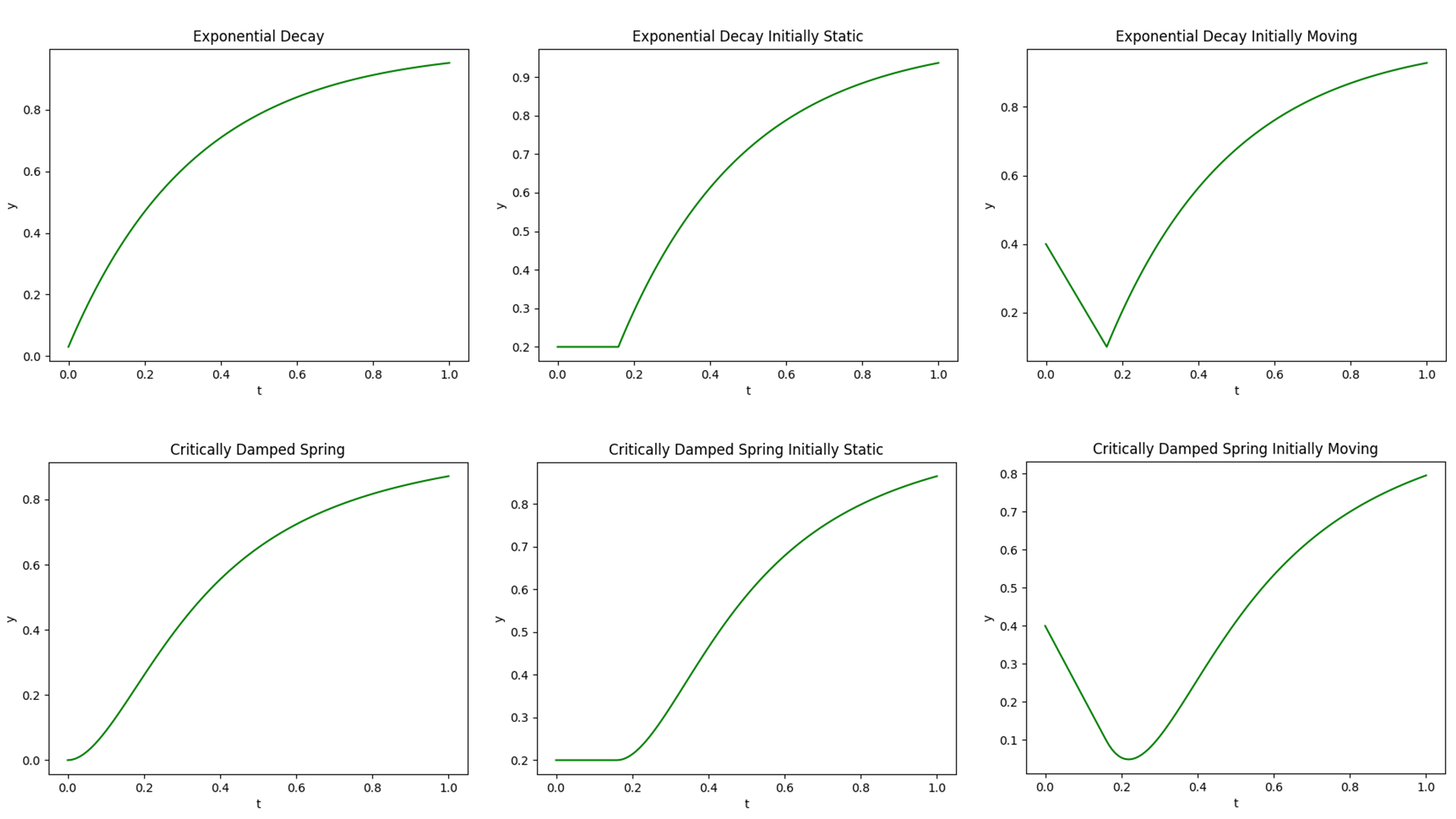}
  \caption[Exponential Decay vs. Critically Damped Spring]{\label{fig:mm:springs} Comparison between exponential decay (top row) and critically damped springs (bottom row) to smooth a value over time. Exponential decay does not have $C^1$ continuity, thus returning sudden changes in value.}
\end{figure}

In the following equations we will work with scalars, but, equations can be extended to work with vectors. Given a target position or direction we could create a smooth trajectory by using an exponential decay function:
\begin{align}
    y_{t + \Delta t} = y_{t} + \beta (\Hat{y} - y_{t}) \Delta t
\end{align}
where $y$ is the value to change over time (time shown in subindices), $\Hat{y}$ is the target value, and $\beta$ modifies the speed to converge to $\Hat{y}$. Although this function produces a smooth change in $y$ as shown in the first row of Figure~\ref{fig:mm:springs}, it will make a sudden change in value whether $y$ was initially already moving or not. In the second row, we can see how a critically damped spring avoids sudden changes in $y$.

Therefore, we use a damped spring model in which the spring's natural length is the desired position $\Hat{y}$ and the end position of the spring is $y$. By combining Hooke's law with an additional force acting against $y$ velocity (damper) we have:
\begin{align}
    m \frac{\partial^2 y}{\partial t^2} = k (\Hat{y} - y) - b \frac{\partial y}{\partial t}
\end{align}
where $m$ is the end point's mass, $k$ is the spring constant, and $b$ controls the amount of drag. If the value of $b$ is small, it may produce overshoot and oscillations, while a large value will have slow convergence. A spring is critically damped when it does not produce oscillations and approaches $\Hat{y}$ at an optimal convergence rate. This occurs when $b^2 = 4mk$ \citep{gems4}, thus:
\begin{align}
   \frac{\partial^2 y}{\partial t^2} = w^2 (\Hat{y} - y) - 2w \frac{\partial y}{\partial t}
\end{align}
where $w = \sqrt{k/m}$ and measures the stiffness of the spring.

\newpage
The critically damped spring can be approximated using numerical integration, but an analytical solution exists \citep{gems4}:
\begin{align} \label{eq:mm:spring_pos}
    y_{t+\Delta t} &= \Hat{y} + ((y_t - \Hat{y}) + (\Dot{y}_t + w(y_t - \Hat{y})) \Delta t) e^{-w \Delta t} \\
    \Dot{y}_{t+\Delta t} &= (\Dot{y}_t - (\Dot{y}_t + w(y_t - \Hat{y}))w \Delta t) e^{-w \Delta t} \label{eq:mm:spring_vel}
\end{align}
where $\Dot{y}_t$ is the velocity at time $t$.

Finally, we can interpret the joystick (2 degrees of freedom) of a controller as the target direction or map some keys from the keyboard to specific directions (e.g., \emph{w} to forward direction). Then, use Eq.~\ref{eq:mm:spring_pos} to smoothly converge to the target direction. The only requirement is to keep the velocity value $\Dot{y}_t$. Assuming constant velocity, we can directly predict directions by computing the position of the spring at different $\Delta t$. The same method could be used to predict future positions in a path; the target position can be the next keypoint in the path or some clamped target in the middle to control the maximum velocity.

Predicting future positions when using a joystick or keyboard is slightly different. The idea is to interpret Eq.~\ref{eq:mm:spring_pos} as if the spring's position was the character's velocity. Then integrating the equation, we can get the position. This approach is used because when using a joystick, we can interpret it as a target velocity, but we do not have a target position.

\subsection{Procedural touch-ups} \label{sec:mm:procedural_touch_ups}
% adjustment; clamping; etc.
In the last section, we discussed critically damped springs to create the trajectory features for the query vector. Then, Motion Matching finds the best sequence of poses that follow the trajectory while keeping a similar pose by minimizing the pose features. A sequence of poses may better match our target trajectory, but another may be chosen to prevent considerable pose changes. Providing weights for the different features in the query vector $\mathbf{\Hat{z}}$ (see Eq.~\ref{eq:mm:search_weights}) helps to adjust this quality vs. responsiveness trade-off. However, even if we set the weights for the current pose features to zero, it is impossible to always find a perfect match with the target trajectory since we are constrained by the sequences of poses available in the animation database.

Ideally, the simulated character and the virtual root joint (virtual character) should be aligned most of the time. But if this is not the case, some procedural touch-ups can help minimize the distance between them. Firstly, we reduce this problem by slightly moving the virtual root joint proportionally towards the simulated character to its velocity, thus minimizing the adjustment when the character moves slowly to avoid users noticing the foot sliding introduced. This method can also be applied to the orientation of the character.

\newpage
Secondly, we provide a position accuracy parameter $\alpha$ to ensure that the position of the virtual root joint $\mathbf{p}$ does not deviate more than $\alpha$ with respect to the simulated character's position $\mathbf{\Hat{p}}$. The corrected position of the virtual root joint $\mathbf{p'}$ is computed as follows:
\[
\mathbf{p'}=
\begin{cases}
    \mathbf{\hat{p}} + \alpha \frac{\mathbf{p} - \mathbf{\hat{p}}}{\lVert \mathbf{p} - \mathbf{\hat{p}} \rVert}, & \lVert \mathbf{p} - \mathbf{\hat{p}} \rVert > \alpha \\
    \mathbf{p}, & \text{otherwise}
\end{cases}
\]

\section{Inertialization} \label{sec:mm:inertialization}
\begin{figure}[h]
  \centering
  \includegraphics[width=0.9\linewidth]{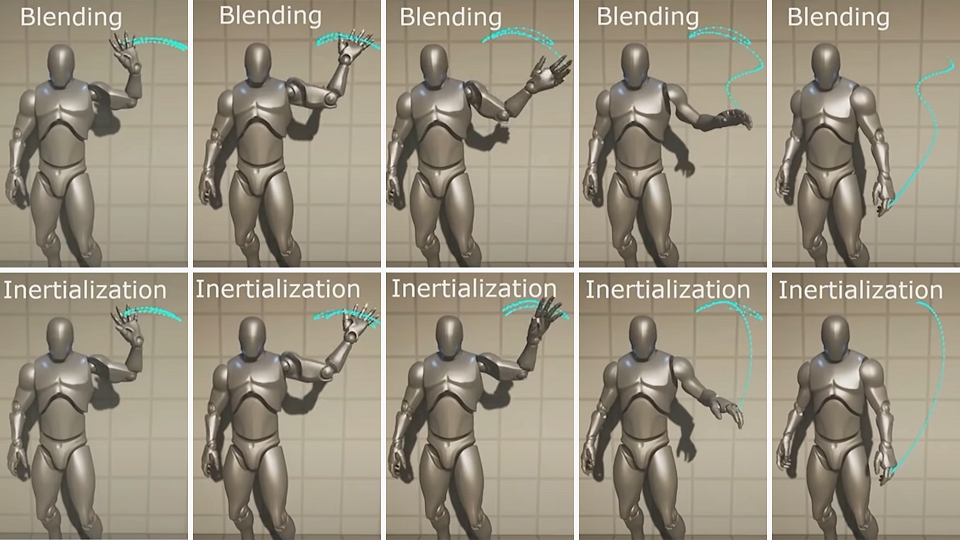}
  \caption[Traditional Blending vs. Inertialization]{\label{fig:mm:inertialization} Comparison between traditional blending (top row) and Inertialization (bottom row). Note how traditional blending keeps waving the hand during the whole motion due to interpolation, while Inertialization tends to produce a natural movement. Source: \citep{bollo2017}.}
\end{figure}

A discontinuity is introduced after performing a Motion Matching search if the returned pose is not consecutive with the current one. Therefore, poses should be blended to avoid significant changes. Traditional blending techniques interpolate by evaluating the two poses during the transition. \citet{bollo2017} presented Inertialize blending, developed for Microsoft's game \emph{Gears of War 4}. This technique only evaluates one pose by decaying the offset relating the target pose with the source pose. Not only is the performance improved, but it also produces a more natural-looking motion than traditional techniques. For instance, as shown in Figure~\ref{fig:mm:inertialization}, if the character is waving the hand and we change to an idle animation, with traditional blending, the hand will continuously wave during the transition due to the interpolation. However, Inertialization smoothly changes to the idle pose without waving.

\begin{figure}[h]
  \centering
  \includegraphics[width=1.0\linewidth]{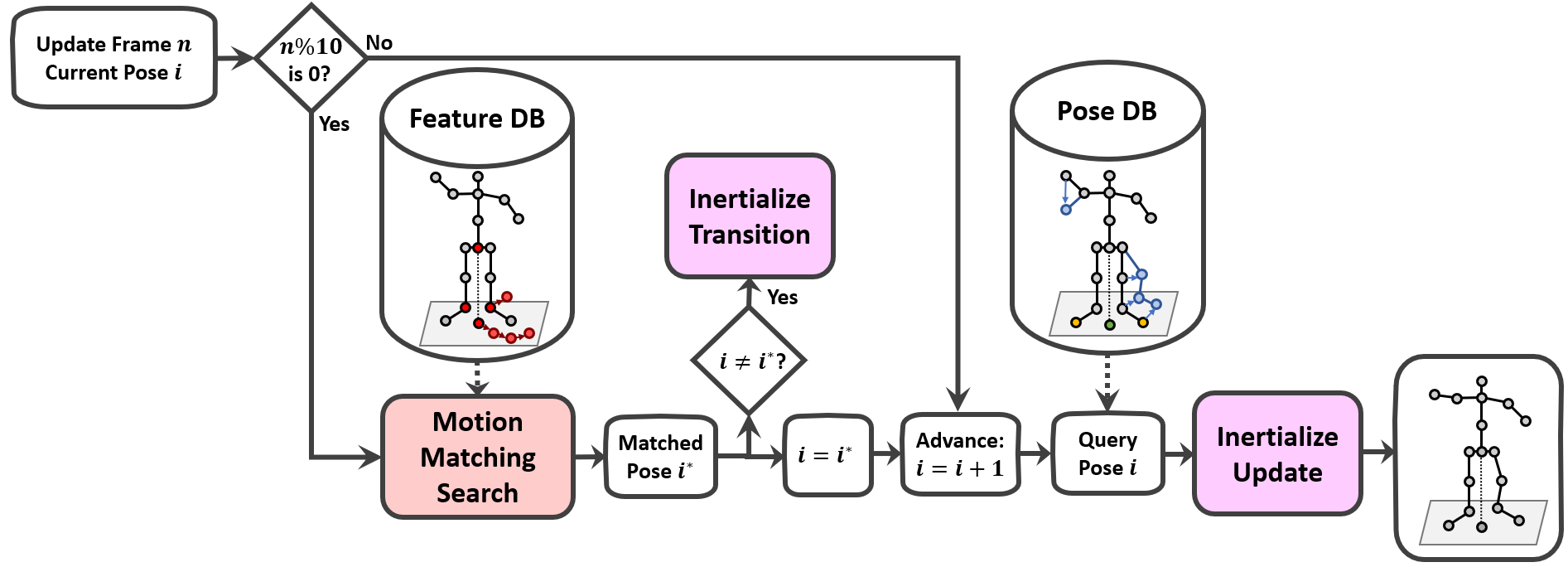}
  \caption[Inertialization Transition and Update]{\label{fig:mm:inertialization_diagram} An inertialize transition is set if there is a change in pose after a Motion Matching search. The transition stores the offset between the source and target poses. The offset decays in every inertialize update.}
\end{figure}

Inertialization consists of two methods (see Figure~\ref{fig:mm:inertialization_diagram}): the transition method stores the offsets between the local rotations from the target pose to the source pose, and the update method decays the offsets. A transition is introduced every time we change to a non-consecutive pose. Given a joint, we define its target local rotation $\mathbf{t^r}$, target local angular velocity $\mathbf{t^w}$, source local rotation $\mathbf{s^r}$ and source local angular velocity $\mathbf{s^w}$. The transition is introduced by modifying the current rotation offset $\mathbf{d^r}$ and angular velocity offset $\mathbf{d^w}$:
\begin{align}
    (\mathbf{d^r})' &= (\mathbf{t^r})^{-1} \mathbf{s^r} \mathbf{d^r} \\
    (\mathbf{d^w})' &= (\mathbf{s^w} + \mathbf{d^w}) - \mathbf{t^w}
\end{align}
where $(\mathbf{d^r})'$ and $(\mathbf{d^w})'$ are the offsets after the transition, rotations are represented by quaternions and angular velocities by axis-angle rotation vectors (with the angle encoded as the length of the axis vector).

The update method is applied to every frame and decays the offset using, for example, the critically damped spring Eq.~\ref{eq:mm:spring_pos}-\ref{eq:mm:spring_vel} with desired position and velocity equal to zero. Thus, after decaying the offset, the inertialized rotation $(\mathbf{y^r})'$ and angular velocity $(\mathbf{y^w})'$ is computed as follows:
\begin{align}
    (\mathbf{y^r})' &= \mathbf{t^r}  \mathbf{d^r} \\
    (\mathbf{y^w})' &= \mathbf{t^w} + \mathbf{d^w}
\end{align}

\section{Implementation}
In the previous sections, we introduced theoretical and practical concepts to develop an animation system based on Motion Matching. However, details about the developed application were omitted to separate the explanation of the algorithm from the implementation. The purpose of this section is to explain some critical implementation details and features developed for our application.

\subsection{Code architecture}
One of the purposes of this work is to create a package that can be easily incorporated into any Unity project. Also, we would like to use the application to continue researching and improving some parts of the system, therefore, it should be modular and allow for easy changes in the different sub-systems.

\begin{figure}[h]
  \centering
  \includegraphics[width=1.0\linewidth]{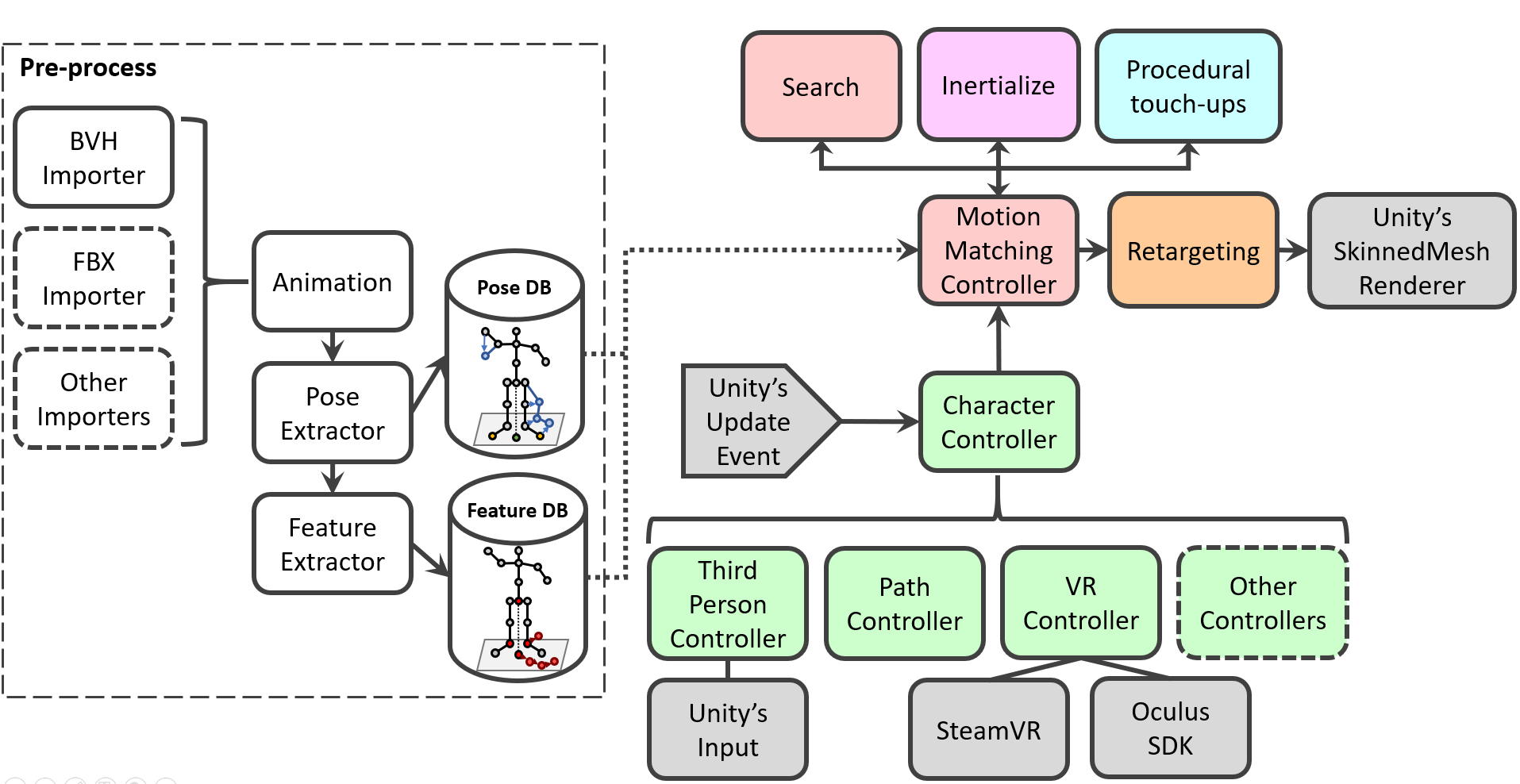}
  \caption[Code Architecture]{\label{fig:mm:code_architecture} Application structure. The pose and feature databases are pre-processed. Every frame, Unity sends an Update event to the character controller, an abstract class allowing for easy implementation of new controllers. Next, the Motion Matching controller class manages the search, inertialization and other procedural touch-ups. The final pose is retargeted and sent to Unity's rendering system. Classes outlined with discontinuous lines are not yet implemented.}
\end{figure}

The architecture of the code is shown in Figure~\ref{fig:mm:code_architecture}. First, a pre-processing step takes an animation file and transforms it into a standardized \emph{Animation} object. The \emph{Animation} class abstracts the actual file in which the animation was saved. For this work, we implemented a BVH importer (see Section~\ref{sec:mm:bvh}), but other importers could be added without modifying the rest of the system. Then, the pose and feature database are created. These databases and other information, such as the animation skeleton, are serialized and stored in a file.  

When the application starts, Unity triggers an \emph{Update} event for every frame, which is received by the abstract character controller class. This class is overridden with different controllers: a third-person controller allows to move a virtual character with a joystick or keyboard; a path controller allows to place keypoints in the scene that the virtual character will follow, and users can also indicate the velocity between keypoints; a VR controller explained in Chapter~\ref{chap:vr}; other controllers could be easily added by overriding the abstract class. Different features can be defined for each type of character controller; this is further explained in Section~\ref{sec:mm:mm_data}.

Once the character controller has updated its state and computed the trajectory features, it sends all information to the Motion Matching controller class: the main class in the animation engine responsible for executing Motion Matching. Trajectory features are received as an array of \emph{floats} following the features definition (see Section~\ref{sec:mm:mm_data}). Therefore, it does not need to know which specific type of character controller is being used and does not need to be re-implemented every time the features definition is changed. The search is implemented in a separate class so different algorithms can be easily incorporated. The following classes also contribute to the final pose: the \emph{Inertialize} class handles the state of the offsets and provides convenient \emph{Transition} and \emph{Update} methods; other classes help to procedurally adjust the final pose and position of the virtual character.

Finally, when the pose is ready, it is sent to the retargeting algorithm, which also transforms the skeleton so Unity's rendering system can understand it. Using the Unity's \emph{SkinnedMeshRenderer} component and the previous retargeting, we can take full advantage of the rendering facilities and animate any humanoid-like virtual character.

\newpage
\subsection{Animation and features definition} \label{sec:mm:mm_data}
\begin{figure}[h!]
  \centering
  \includegraphics[width=1.0\linewidth]{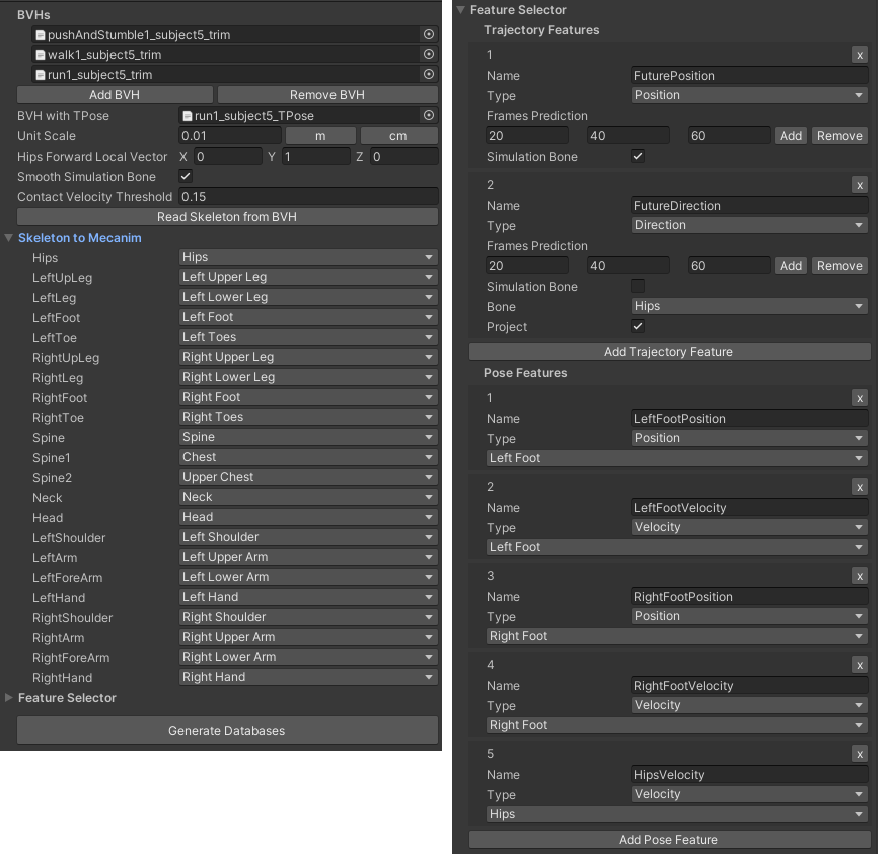}
  \caption[Animation and features definition]{\label{fig:mm:mmdata} User interface to define the animation files and features used by Motion Matching. On the left, a list of BVHs and import options are defined, then the user specifies the mapping from the animation skeleton to the humanoid convention for retargeting. On the right, users can determine the trajectory and pose features. Finally, databases are computed once the \emph{Generate Databases} button is pressed.}
\end{figure}

Prior to executing the system, users must define the animation files and features that will be used by Motion Matching. For this, we created a user interface as shown in Figure~\ref{fig:mm:mmdata}. Settings are stored into a configuration file used by the Motion Matching and character controllers. The user interface is divided into three parts:
\begin{itemize}
    \item \textbf{BVH and import settings}. Users specify one or more BVH to create the animation database and one BVH containing a T-Pose in the first frame. If the skeleton among all BVHs is not the same, the UI will display an error. Next, users define the BVH scale, the hip joint's default forward direction, and some other options for the pose database creation.
    \item \textbf{Skeleton mapping}. In Section~\ref{sec:mm:retargeting} we use a general humanoid skeleton convention to allow retargeting between virtual characters. The UI reads the skeleton defined in the BVH and displays a list so users can specify the mapping of each joint to its correspondent in the convention.
    \item \textbf{Feature definition}. Although in this work we presented those features used for locomotion, the UI allows to customize both trajectory and pose features. Several options can be selected, such as the type (position, direction, velocity), the joint, and the samples in the future (or past) in the case of trajectory features.
\end{itemize}
Once all settings are set, the user presses the \emph{Generate Databases} button, and the system automatically creates the pose and feature databases and stores them in a file.

\newpage
\subsection{High-performance search in Unity} \label{sec:mm:performance}
One essential aspect of real-time computer animation is performance; nowadays, modern game productions have around 1 millisecond (ms) CPU budget for computing all animations in a scene. The search is the most performance-critical component of Motion Matching. \citet{holden2020} presented their C++ Motion Matching implementation and evaluated the performance. Using the Ubisoft's database \emph{LAFAN1} \citep{harvey2020robust} they were able to search in an average of 0.09\;ms. This section summarizes how we implemented the search in Unity with equivalent performance.  

Unity provides a scripting layer to implement new functionalities with the C\# programming language. To obtain a performant search, we first consider the implementation of the feature database. At first, an Object-oriented programming (OOP) approach would implement the feature database as an array of classes of type \emph{FeatureVector}. Each class would contain all features described in Eq.~\ref{eq:mm:z}. However, since classes are reference-types in C\#, they might not be allocated contiguously in memory, drastically decreasing the search performance due to a high number of \emph{cache misses}. Instead, the feature database should be stored as an array of value-types such as \emph{floats} or structs of \emph{floats}.

Another issue is that Unity uses by default the Mono compiler, which translates C\# code into Intermediate Language (IL) code. This approach is typically enough for most scripts, however, it may not produce acceptable performance when used for compute-intensive algorithms. For instance, in Table~\ref{table:mm:performance} (column \emph{Linear}), we report the results of a linear search implemented with arrays contiguous in memory and compiled with Mono. The same method is tested in three different databases: our custom animation database ($18{,}120$\;poses), Ubisoft's \emph{LAFAN1} ($14{,}867$\;poses) and our custom database for VR avatars ($28{,}271$\;poses). If we look at our custom animation database, this method uses on average 17.36\;ms for every search.

\begin{table}[h]
\centering
\begin{tabular}{| c || c | c | c | c | c |}
 \hline
 \multicolumn{6}{|c|}{\textbf{Search Performance (ms)}} \\
 \hline
 Database & Linear & Linear+Burst & Linear Optimized+Burst & BVH & BVH+Burst \\
 \hline
 Xsens & 17.36 & 0.45 & 0.21 & 0.31 & \textbf{0.03} \\
 \hline
 Ubisoft & 14.26 & 0.84 & 0.48 & 0.60 & \textbf{0.07} \\
 \hline
 VR & 27.07 & 1.58 & 0.91 & 0.39 & \textbf{0.05} \\ 
 \hline
\end{tabular}
 \caption[Search Performance]{\label{table:mm:performance} Search performance comparison for three animation databases. Linear searches can obtain good results when compiled with Burst and optimized for SIMD instructions. The best performance is obtained when using a Bounding Volume Hierarchy compiled with Burst. Numbers are averaged over 100 searches.}
\end{table}

Instead, we can use Unity's Burst compiler that translates IL code to optimized native code. The only restriction is that it uses a C\#-like syntax but does not have all the features. This compiler automatically applies vectorization to some parts of the code; that is, it can combine multiple CPU instructions into one SIMD instruction, thus exploiting data-level parallelism. As shown in the table (column \emph{Linear+Burst}), the linear search compiled with Burst obtains a significant speedup and uses on average 0.45\;ms for every search. We can further improve this number by manually applying vectorization (column \emph{Linear Optimized+Burst}) and obtain an average of 0.21\;ms per search.

Finally, we also implemented a Bounding Volume Hierarchy (BVH) acceleration structure as described by \citet{holden2020}. It uses axis-aligned bounding boxes (AABB) organized in two layers of groups of 64 and 16 consecutive frames. The BVH implementation compiled with Mono already obtains good results (column \emph{BVH}) with an average of 0.31\;ms for every search. When compiled with Burst (column \emph{BVH+Burst}), we obtain the best results: 0.03\;ms per search.

Considering that the search is performed every ten frames, every search takes around 0.03\;ms, and we can distribute searches among frames for different virtual characters; with an animation budget of 1\;ms per frame, we could animate more than 300 virtual characters simultaneously.

\addtocontents{toc}{\protect\newpage} % Avoid index splitting this chapter
\chapter{Virtual Reality Avatars} \label{chap:vr}

Animating self-avatars in VR when using a Head Mounted Display (HMD) is vital to provide embodiment and improve the user experience when it comes to understanding dimensions and distance to objects or other users in the case of collaborative VR. High-end mocap systems provide outstanding animations but at a high cost. Ideally, users would need correct animations with standard VR devices, which typically include three trackers: HMD and hand controllers. Currently, most games align the body orientation with the HMD, thus providing only forward movement, and apply animation blending and time-warping from a reduced set of animations, which causes noticeable mismatches between the movements of the avatar and the user.

In this chapter, we present a system that can correctly extract the user body orientation using a neural network that only requires tracking information from the HMD and hand controllers. Then it uses this direction with velocity and orientation of the three trackers to build a feature vector that serves as input for a Motion Matching algorithm which uses an animation database captured from users walking while wearing a HMD. Our system can provide a large variety of lower body animations while correctly matching the user orientation, allowing us to represent forward movement and stepping in any direction.

The system can be divided into three parts: body orientation prediction, Motion Matching and final pose adjustments. One of the inputs of Motion Matching is the user's trajectory, which is defined in terms of the future positions and the target body orientation. While the future positions can be predicted from the HMD's velocity, estimating the body orientation from the HMD and controllers is not straightforward. Therefore, we designed and trained a neural network to obtain an estimation of the actual body orientation.

Once we have predicted the user's trajectory, we build a query vector and use Motion Matching to search for the sequence of poses that better match our current pose and target trajectory. Our system supports animations such as crouching by letting the Motion Matching algorithm search in different animation databases, depending on the height of the HMD.

Our method focuses on the animation of the lower body, for which tracking information is missing. In contrast, the arms can be animated by applying IK since the end effectors can be located and oriented precisely according to the handheld controllers. This separation of the upper and lower body parts dramatically decreases the dimensionality of the feature query for the Motion Matching search while providing a solution that guarantees the correct positioning of the hands at all times. The following sections describe each part of the method in detail. 

\begin{figure}[h]
  \centering
  \includegraphics[width=0.8\linewidth]{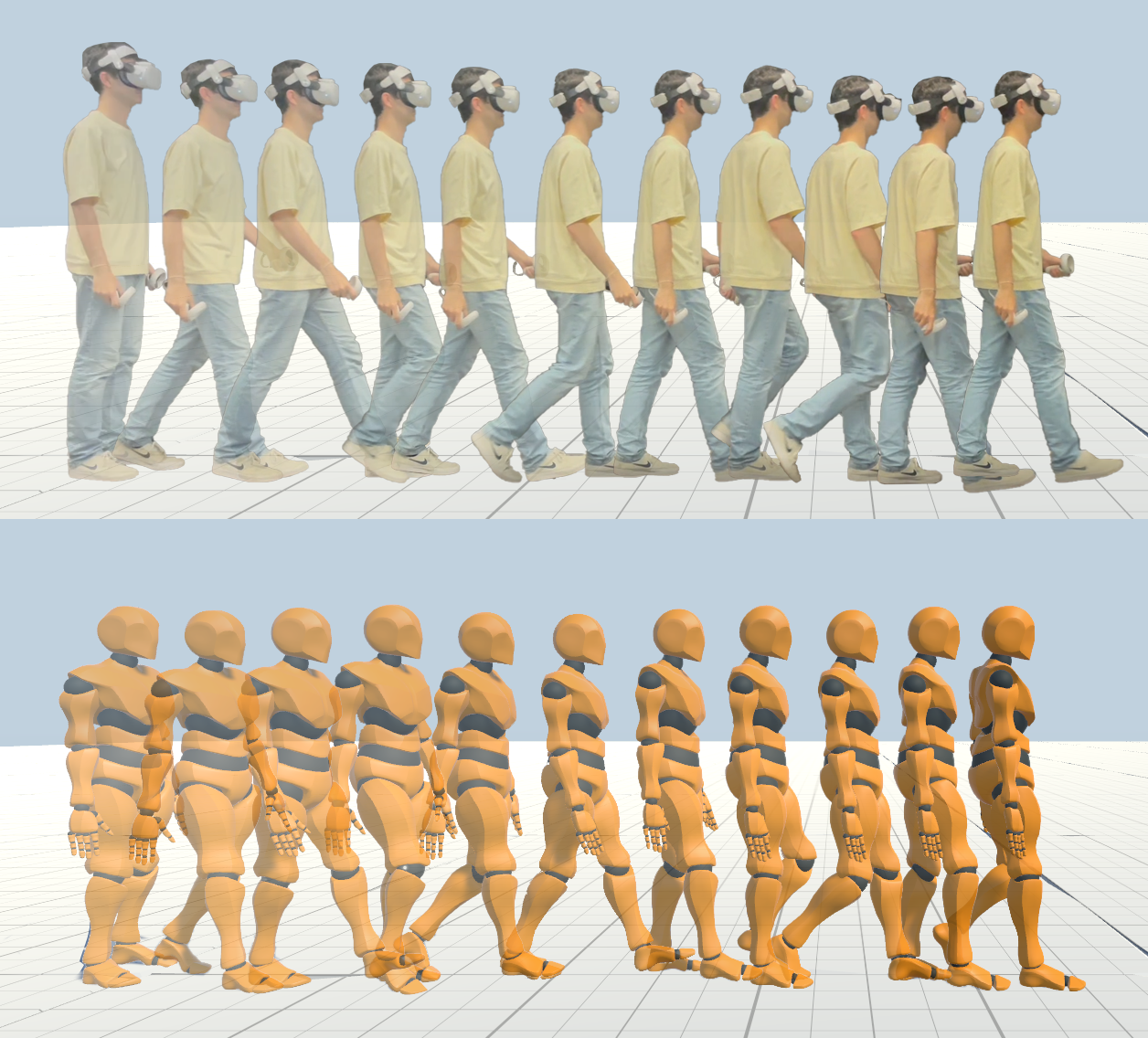}
  \caption[User and VR Avatar]{\label{fig:vr:real_virtual_locomotion} An avatar walking driven by a VR user. The avatar animation is based exclusively on the tracking data provided by the VR headset and two handheld controllers.}
\end{figure}

\newpage
\section{Prediction of the body orientation}
\begin{figure}[h]
  \centering
  \includegraphics[width=0.7\linewidth]{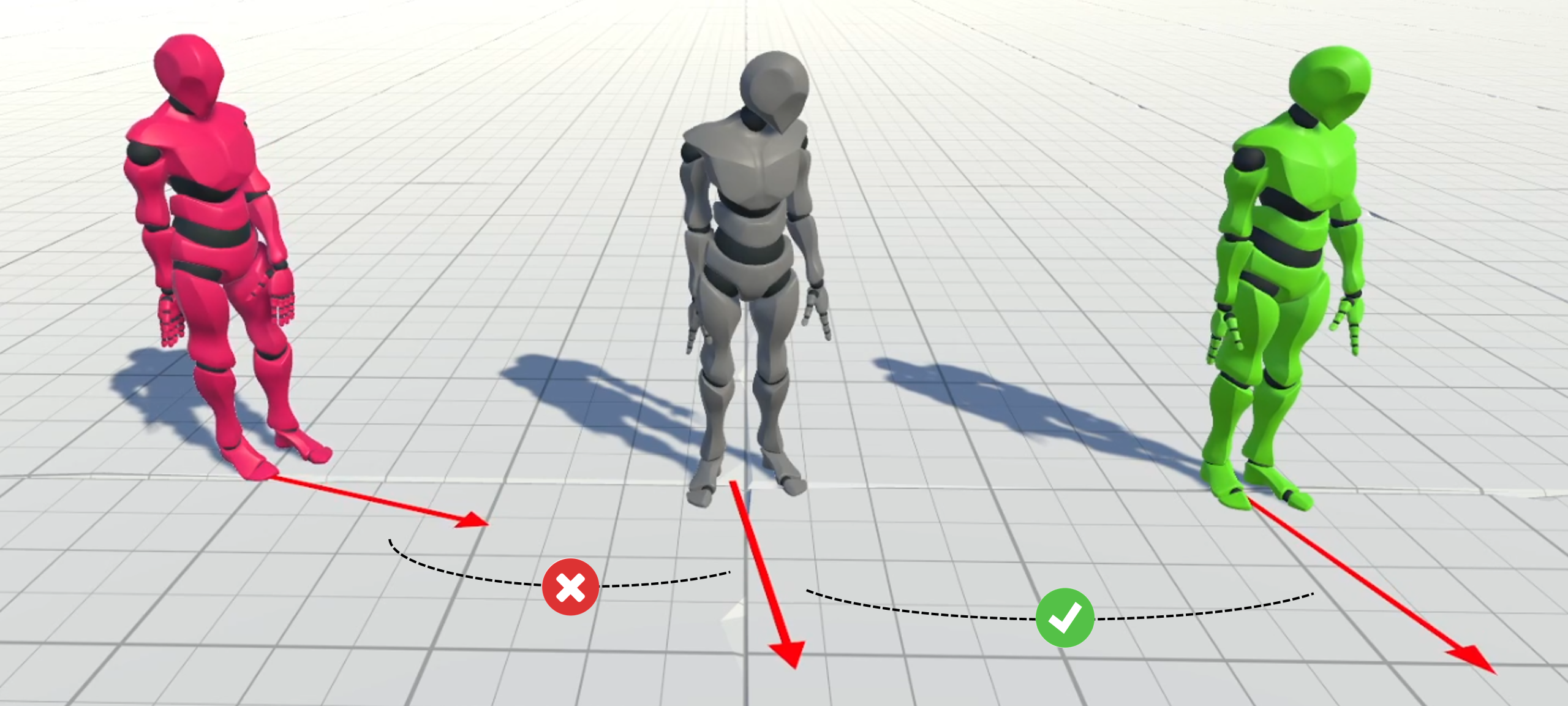}
  \caption[Body direction comparison]{\label{fig:vr:direction_comparison} In the center the ground truth using Xsens with the red vector indicating the correct body orientation. On the left incorrect torso orientation when using the HMD forward vector as body orientation. On the right our neural network prediction.}
\end{figure}

Predicting the body orientation is a common problem in applications using full-body avatars with only one HMD and two controllers. Often, the forward direction of the HMD is used to orient the whole virtual body, as shown in Figure~\ref{fig:vr:direction_comparison}. However, this leads to a significant misalignment between the user's forward direction (and pose) and the avatar. For example, users may be moving their heads to look around (keeping the rest of the body static), but their avatars will rotate the whole body (not just the head) to match the orientation of the HMD.

We cannot directly extract the body's orientation from the HMD and controller. However, we can try to infer it from the information given by these devices. For this purpose, we trained a lightweight feedforward neural network to predict the body orientation from the rotation, velocity and angular velocity of all three devices. Positions are not used so that the method is body-size independent and can be applied to a broader range of users. Our method also takes as an input the previously predicted body orientation to induce temporal continuity.

\subsection{Network input and output}
The input vector for the neural network is constructed using the velocities and rotations of the $k=3$ trackers and the previous predicted orientation. It can be defined as $\mathbf{x} = \left( \mathbf{x^v}, \mathbf{x^w}, \mathbf{x^r}, \mathbf{x^d} \right) \in \mathbb{R}^{k\times12+6}$ where $\mathbf{x^v} \in \mathbb{R}^{k\times3}$ are the 3D velocities, $\mathbf{x^w} \in \mathbb{R}^{k\times3}$ are the 3D angular velocities (axis-angle rotation vector with the angle encoded as the length of the axis vector), $\mathbf{x^r} \in \mathbb{R}^{k\times6}$ are the 3D rotations, and $\mathbf{x^d} \in \mathbb{R}^{6}$ is the previous predicted orientation. Rotations are represented with the 2-axis rotation matrix \citep{zhou2019} to ensure rotation continuity during training. We also normalize each feature of the input by subtracting their mean and dividing by the standard deviation. The output $\mathbf{\Hat{d}} \in \mathbb{R}^{6}$ is the predicted body orientation. 

To create the database, we used an Xsens motion capture system to record the full-body motion of a user while playing different games for SteamVR-based HMDs and Oculus Quest. We also explicitly captured more extreme movements to cover a wide range of poses. In total, we used around half a million poses for the training with around 2.4 hours of motion capture. We use this data to simulate the trackers' information. We assume fixed offsets between the head, left and right wrist joints and the corresponding trackers. Then, we compute their velocities, angular velocities and rotations for each pose. We represent the body's orientation with the orientation of the virtual root joint computed in Section \ref{sec:mm:pose_database}. To facilitate training, we represent all features with respect to a local coordinate system defined by the following three axes: the projection of the HMD's forward direction onto the floor plane, the vertical world vector, and their cross product. Local space guarantees independent prediction from user's orientation.

\subsection{Network architecture and training} \label{sec:vr:network_training}
\begin{figure}[h]
  \centering
  \includegraphics[width=0.9\linewidth]{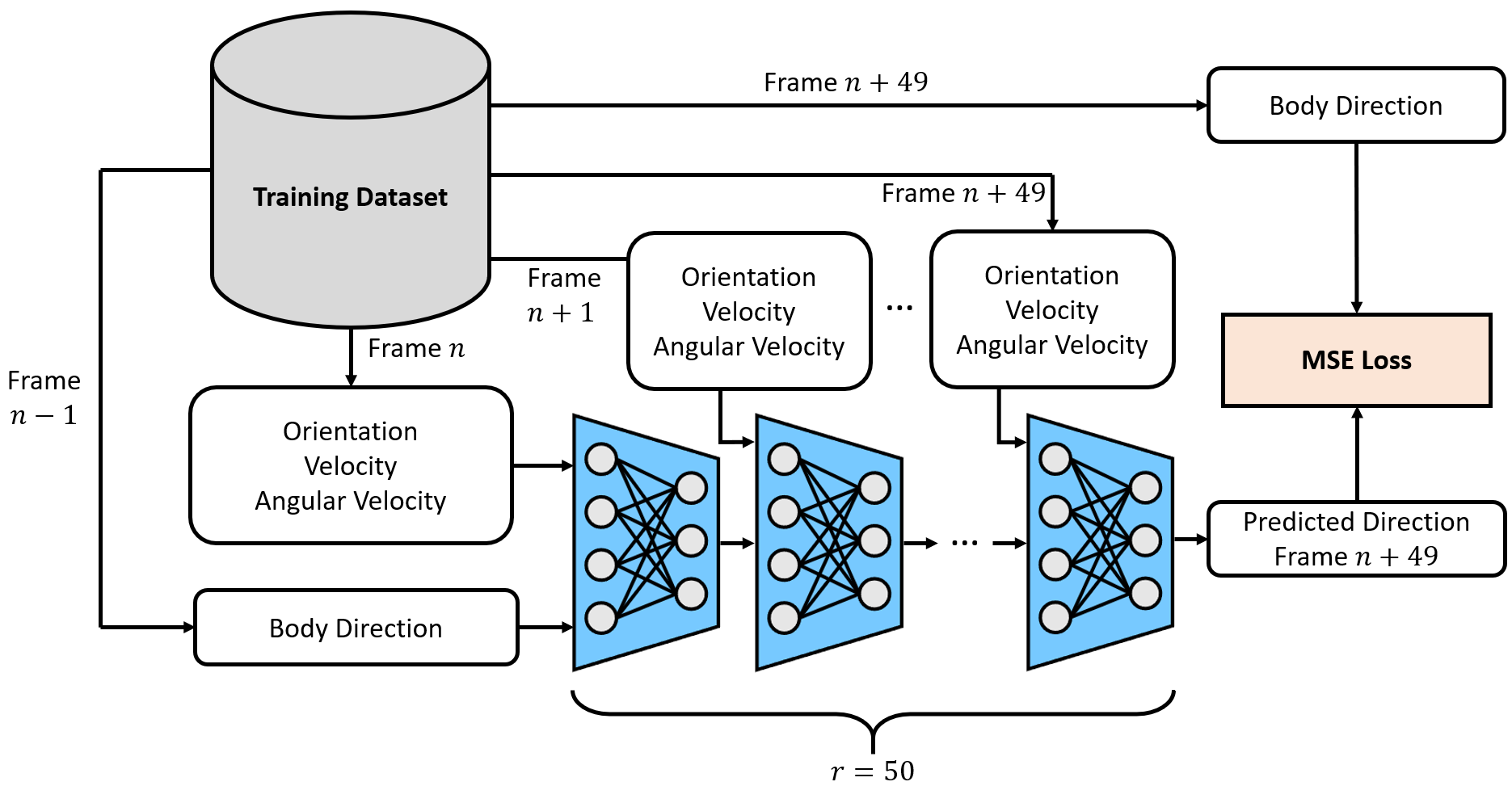}
  \caption[Orientation Prediction Training]{\label{fig:vr:network_training} Training procedure for the body orientation predictor neural network. The orientation is predicted $r$ times before computing the MSE Loss.}
\end{figure}

We used a simple feedforward neural network with 2 hidden layers with 32 units each and ReLU activation functions. The training is performed as usual with feedforward neural networks. However, predicting the orientation directly from the ground truth data would not match the real usage scenario of the network, and therefore, the network would not be learning how to predict the next orientation based on the previously predicted one. Instead, for every element in a training batch, we iteratively predict the orientation $r$ times (e.g., $r=50$) as shown in Figure~\ref{fig:vr:network_training}. Then, we compute the MSE loss by comparing the final predicted body orientation $\mathbf{\Hat{d}}$ with the ground truth orientation $\mathbf{d^*}$ after $r$ frames.

\section{Motion Matching for VR}
Chapter~\ref{chap:mm} contains an in-depth explanation of Motion Matching for animating virtual characters. This section focuses on the specific details needed to use Motion Matching for VR avatars. 

\subsection{Trajectory features for VR}

Since we are using Motion Matching for locomotion, the pose and feature database definition remains unchanged. However, when creating the trajectory features, we need to get the information from the HMD and the controllers. First, we obtain the HMD velocity $\mathbf{v}$, although directly using it could lead to undesirable noise and discontinuities. Instead, the velocity is smoothed using a critically damped spring as in Eq.~\ref{eq:mm:spring_pos}. Similarly, we do not directly use the predicted body orientation $\mathbf{\Hat{d}}$, but a smoothed version.

Formally, to create a query vector $\mathbf{\Hat{z}}$ we retrieve the current feature vector $\mathbf{z}$ from the feature database, and predict the trajectory of the user from $\mathbf{v}$ and the orientation $\mathbf{\Hat{d}}$:
\begin{align}
    \mathbf{\Hat{z}}^\mathbf{v} &= \mathbf{z}^\mathbf{v} \\
    \mathbf{\Hat{z}}^\mathbf{l} &= \mathbf{z}^\mathbf{l} \\
    \mathbf{\Hat{z}}^\mathbf{p} &= ( \mathbf{\Hat{p}} + \frac{1}{3} \mathbf{v}, \ \mathbf{\Hat{p}} + \frac{2}{3} \mathbf{v}, \ \mathbf{\Hat{p}} + \mathbf{v} ) \\
    \mathbf{\Hat{z}}^\mathbf{d} &= ( \mathbf{d}_{20\Delta t}, \ \mathbf{d}_{40\Delta t}, \ \mathbf{d}_{60\Delta t} )
\end{align}
where $\mathbf{\Hat{z}}^\mathbf{p}$ and $\mathbf{\Hat{z}}^\mathbf{d}$ contain the future predictions at $0.33$, $0.66$ and $1.00$ seconds assuming the application runs at 60 frames per second. The orientation $\mathbf{d}$ is estimated with Eq.~\ref{eq:mm:spring_pos} at 20, 40 and 60 frames ahead, and the target position of the body $\mathbf{\Hat{p}}$ is computed by projecting the center of the head on the ground floor (the center is an estimate of the most stable point under rotations of the head). All values are local to the virtual root joint.

\subsection{Position accuracy} \label{sec:vr:pos_accuracy}
One of the limitations of Motion Matching is the drift between the desired and actual position and direction of the character. The search tries to find a sequence of poses that follows the target trajectory while avoiding significant changes in the pose. A sequence of poses may better match our target trajectory, but another may be chosen to prevent considerable pose changes. This issue was discussed in Section~\ref{sec:mm:procedural_touch_ups} in which we provide a position accuracy parameter that ensures that the position of the virtual root joint $\mathbf{p}$ does not deviate more than $\alpha$ with respect to the target position $\mathbf{\Hat{p}}$.

The position accuracy parameter is usually unnecessary when animating a character in a video game from a third-person view. However, it can be problematic when animating a self-avatar in VR, where a correct alignment between the virtual character and the user is always needed. If positional accuracy is prioritized, users can use a low $\alpha$ (e.g., 10\,cm) to reduce the positional misalignment that, in the case of a self-avatar in VR, could lead to a reduction in the Sense of Embodiment. On the contrary, a larger value will leave more freedom for Motion Matching to provide higher quality motions at the expense of some positional drift, which may not be significant when animating other users' virtual avatars for collaborative VR. Therefore, $\alpha$ can be adjusted depending on the application's requirements and the user preference.  

We captured specific motions typically found in VR to reduce further the position accuracy issue: slow movements (users tend to walk carefully in VR), with different velocities for the same movement, and sudden changes in velocity direction and torso orientation. A good alignment between the animation database and the user movements is critical to minimize deviation when searching for trajectories.

\subsection{Non-upright motions}
One crucial aspect of keeping users immersed in VR is synchronizing the leg movements of the avatar with the users. \citet{Ponton2022} suggests that having the bending of the virtual legs synchronized with the user's legs positively affects the Sense of Embodiment in VR. Consequently, we provide a way to control the height of the virtual avatar while maintaining fast Motion Matching searches and motion continuity.

In addition to the standard locomotion database, we captured multiple locomotion databases with different levels of knee bend: from a slight bend to having the legs completely bent or walking on tip-toes. Then, the user's height is represented with a normalized value computed as the ratio between the current HMD's height and the one calculated during a calibration step (at the beginning of the execution, the user is asked to press a button while standing up). Each database has an assigned range of height ratios, and thus, in real-time, we can select the proper database only by querying the HMD's height. 

\begin{figure}[h]
  \centering
  \includegraphics[width=1.0\linewidth]{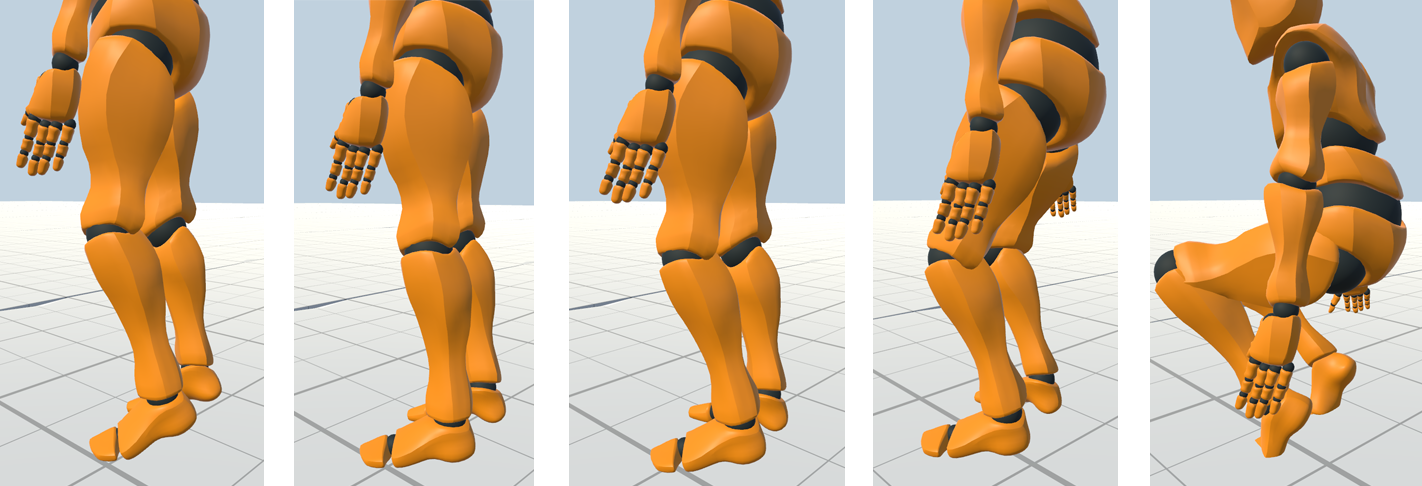}
  \caption[Different leg bending]{\label{fig:vr:leg_bending} Examples of poses with different leg bends based on the HMD height. From left to right: tip-toes, normal length, small knee bend, medium knee bend, and crouching pose.}
\end{figure}

Every time we change the database, we trigger a Motion Matching search. Despite changing the database for the search, we will obtain a similar pose by maintaining the current pose features in the query vector (e.g., the local position of the feet). For instance, if the right foot is ahead of the left one, the new search will try to find a pose with the same foot configuration in the new database. The query vector $\mathbf{\Hat{z}}$ is computed as usual, and the final result is inertialized to blend significant changes in pose. The result for different databases can be seen in Figure \ref{fig:vr:leg_bending}.

\section{Final pose adjustments}
The upper body is not considered for the Motion Matching algorithm to avoid increasing the dimensionality of the feature vector and focus instead on lower body locomotion, for which no tracking data is available in consumer-grade VR. The result of the Motion Matching search is thus a pose from the locomotion animation database. In order to obtain the upper body pose for the arms, we can use the hand controllers as end effectors for an Inverse Kinematics (IK) algorithm. This solution is fast to compute and provides a good solution for the user to interact with the environment in VR.

Suppose the animation database does not contain enough variation in velocities and trajectories. In that case, the search will constantly return motions that slightly deviate from the target trajectory, which, together with the method to avoid positional error, may cause considerable foot sliding. While one possible solution would be to add even more motions to the animation database, this is not always possible due to computation or memory requirements. Therefore we propose to apply foot lock to improve the final result.

Foot lock first checks whether the contact information $\mathbf{y^c}$ returned by the pose vector is set to \emph{true}. If that is the case, inertialization is used to blend the current foot position and the locked one (this is needed because the actual position may not coincide with the pose vector because of the blending). Once the velocity of the foot after inertialization is close to zero, it has reached its final position and should be locked. The lock is applied with an analytical two-joint IK solver for the leg, which uses the pose returned by Motion Matching as a starting point to avoid sudden changes. In the following poses, if $\mathbf{y^c}$ is set to \emph{false} or the distance between the foot's position and the locked position is too far, the foot is unlocked, and inertialization is applied to ensure a smooth transition.
\chapter{Discussion and Results} \label{chap:results}
We have implemented the proposed method for animating virtual characters and VR avatars using the game engine Unity 2021.2.13f1 and PyTorch 1.11.0 \citep{pytorch2019}. VR avatars were tested on Oculus Quest 1 and 2 (standalone headsets), and on a HTC Vive Pro driven by a PC equipped with an Intel Core i7-8700k CPU, 32GB of RAM and an NVIDIA GeForce GTX 1070 GPU. We used FinalIK from \citet{FinalIK} as IK solver to animate the upper body pose. This chapter is structured in two sections. 

Firstly, we present our system working with different animation databases and character controllers. We have not included a performance evaluation in this section since it was already studied in Section~\ref{sec:mm:performance}. Secondly, we show the results of our method for animating VR avatars. We compare it with current solutions for consumer-grade VR used in video games and applications. We present how the system can be easily used in non-locomotion applications, study the effect of different parameters on the final animation, and analyze the body orientation prediction. Our method for animating VR avatars has been submitted to the 21st annual ACM SIGGRAPH / Eurographics Symposium on Computer Animation (SCA 2022).

\section{Motion Matching}

\subsection{Animation database}
\begin{figure}[h]
  \centering
  \includegraphics[width=1.0\linewidth]{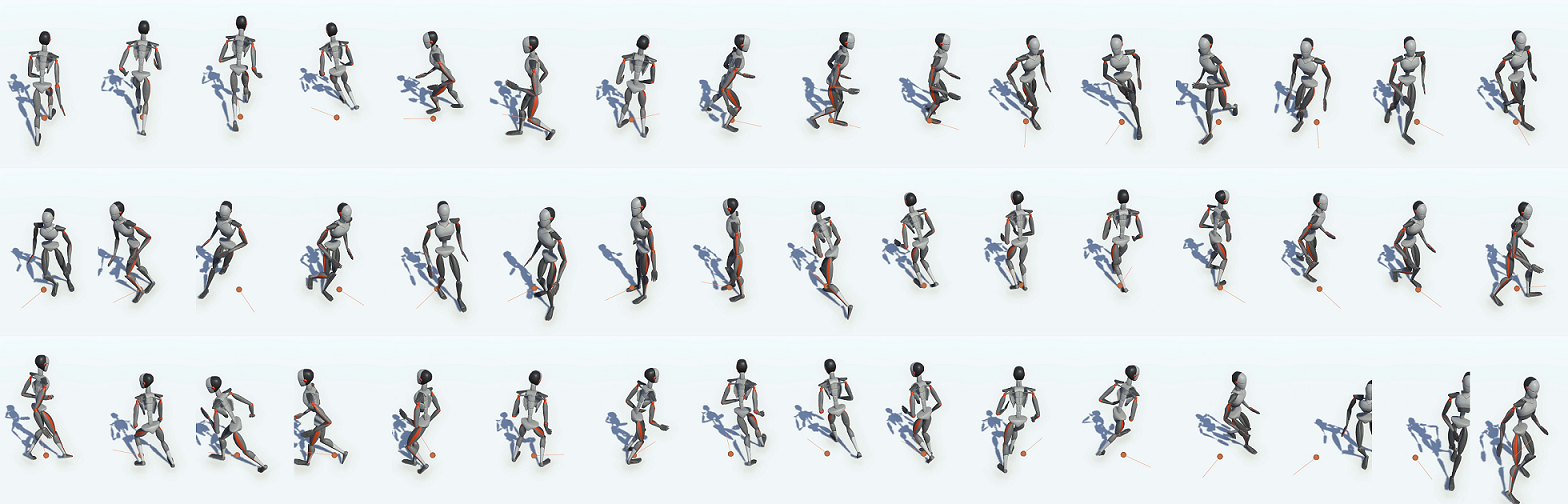}
  \caption[Virtual character animated with our database]{\label{fig:res:jl_mm} Virtual character animated with our motion-captured database. In orange, the position and direction of the simulated character (code-driven by user input).}
\end{figure}

In this section, we show results using two different animation databases for Motion Matching. Using the third-person character controller and a joystick, Figure~\ref{fig:res:jl_mm} shows the Xsens virtual character animated with our custom animation database made of 18,120 poses captured at 60 frames per second ($\sim 5$ minutes). The animation database was recorded as explained in Section~\ref{sec:mm:mocap} and is used directly with our Motion Matching system. No manual labeling or cleaning procedure was needed. The system can produce high-quality animations with a rich set of transitions between poses without creating any state machine. 

\begin{figure}[h]
  \centering
  \includegraphics[width=1.0\linewidth]{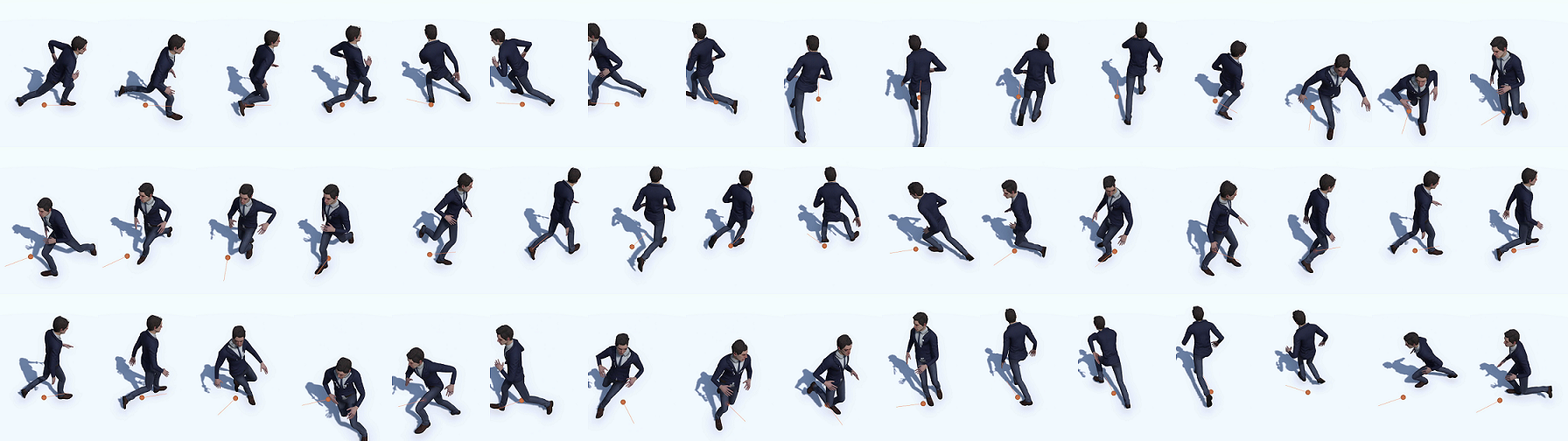}
  \caption[Virtual character animated with Ubisoft's database]{\label{fig:res:ubisoft_mm} Virtual character animated with Ubisoft's database \citep{harvey2020robust}. In orange, the position and direction of the simulated character (code-driven by user input).}
\end{figure}

We also tried the database created by Ubisoft \citep{harvey2020robust} as shown in Figure~\ref{fig:res:ubisoft_mm}. Our implementation retained the high-quality animations, and the virtual character was animated with no extra effort.

\subsection{Character controller}
\begin{figure}[h!]
  \centering
  \includegraphics[width=1.0\linewidth]{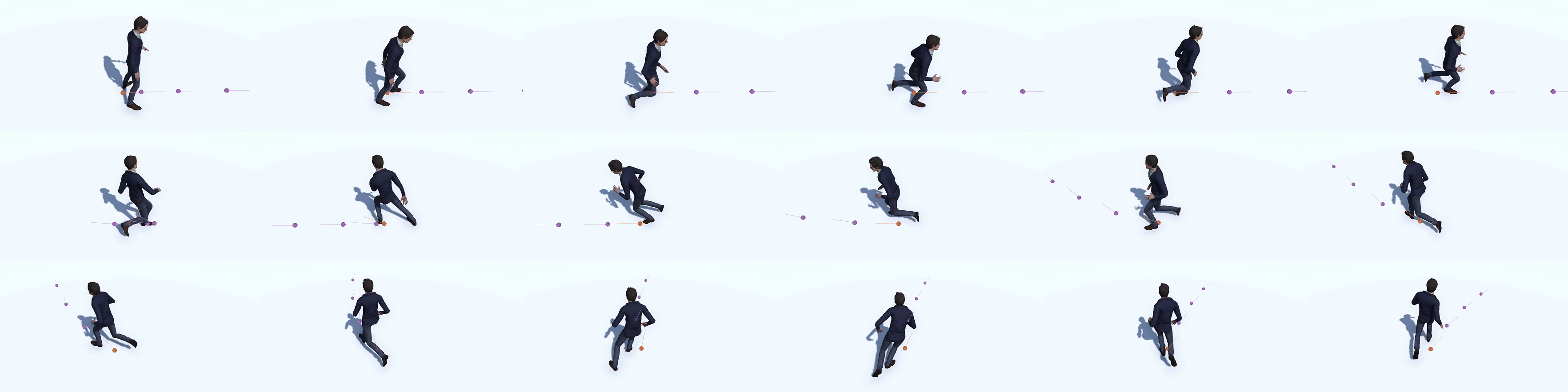}
  \caption[Third-person character controller results]{\label{fig:res:3rd_controller} A virtual character is driven by the third-person character controller. In orange, the position and direction of the simulated character. In purple, the predicted future positions and directions.}
\end{figure}

\begin{figure}[h!]
  \centering
  \includegraphics[width=1.0\linewidth]{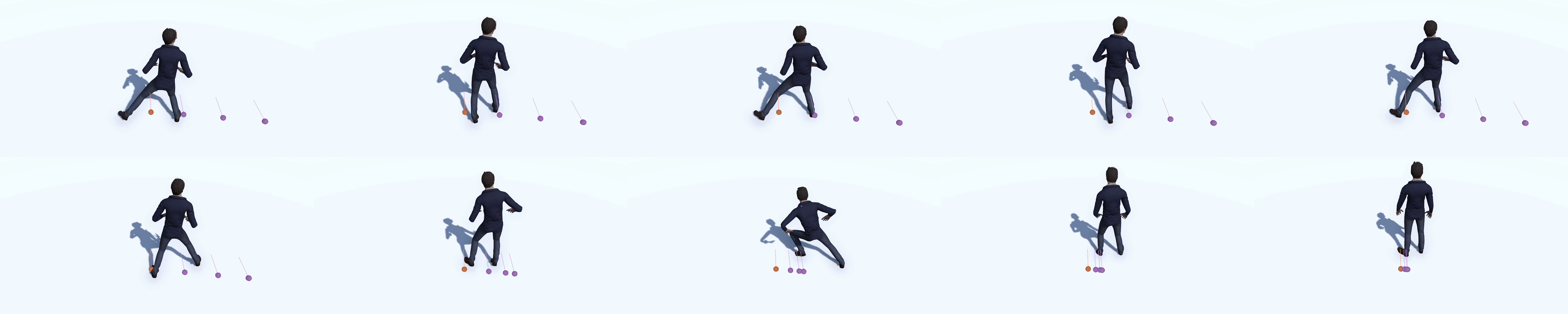}
  \caption[Third-person character controller side stepping]{\label{fig:res:side_stepping} A virtual character is driven by the third-person character controller. In orange, the position and direction of the simulated character. In purple, the predicted future positions and directions. Notice that the direction is fixed, so the virtual avatar moves side-stepping.}
\end{figure}

\begin{figure}[h!]
  \centering
  \includegraphics[width=1.0\linewidth]{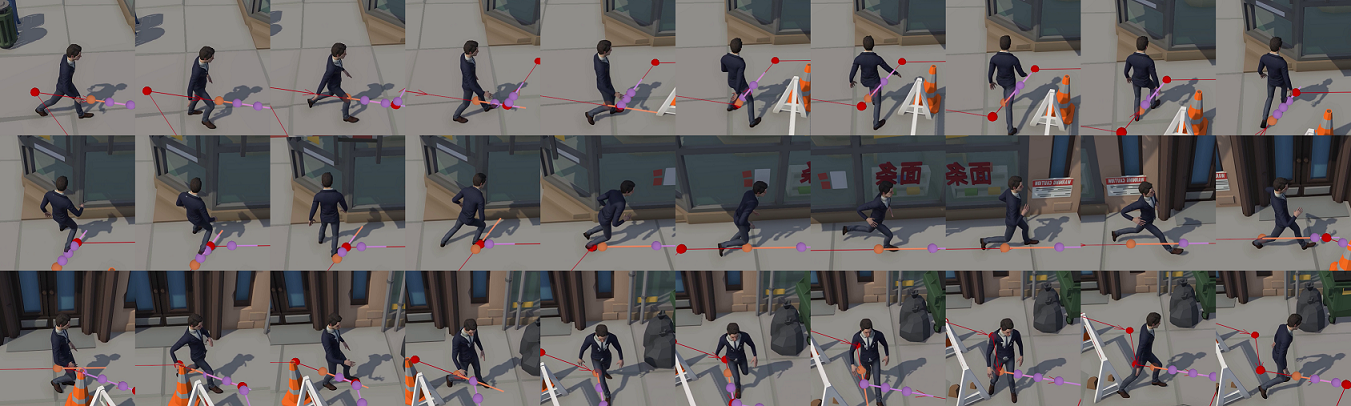}
  \caption[Path character controller results]{\label{fig:res:path} A virtual character is driven by the path character controller. In red, the path. In orange, the position and direction of the simulated character. In purple, the predicted future positions and directions. Some segments of the path have different target velocities.}
\end{figure}

Our system can be used in a wide range of applications. Apart from VR avatars reviewed in Section~\ref{sec:res:vr_avatars}, we have developed two character controllers to demonstrate the versatility of the system. Firstly, a third-person character controller driven by a joystick or keyboard. It predicts the future position and directions with a critically damped spring, as shown in Figure~\ref{fig:res:3rd_controller} the virtual character is capable of following the generated trajectories while creating continuous animations. The direction of the character can also be fixed as in Figure~\ref{fig:res:side_stepping}. Secondly, a path character controller allows to define keypoints in the scene and joins them, creating a path. Motion Matching smoothly finds trajectories that make the virtual character follow the path as shown in Figure~\ref{fig:res:path}. 

\section{VR avatars} \label{sec:res:vr_avatars}

\subsection{Comparison}
In this section, we highlight the main advantages of our method compared to standard solutions found in current VR applications. We focus on four categories of motions that are typical movements in VR. 

\begin{figure}[h]
  \centering
  \includegraphics[width=1.0\linewidth]{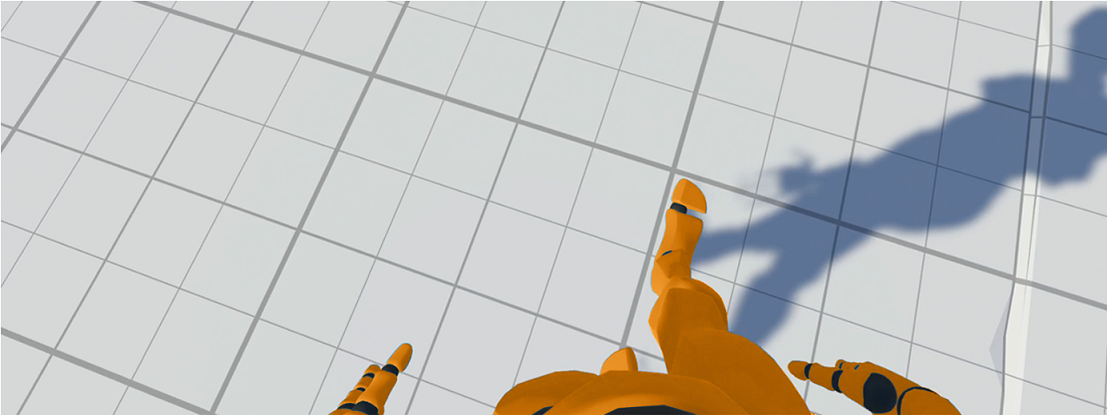}
  \caption[VR Avatar seen from a first-person view]{\label{fig:res:dancing} VR avatar seen from a first-person view.}
\end{figure}

\paragraph*{Walking}
When the user is physically walking, most VR applications animating full-body avatars drag the character, procedurally generate feet position and apply IK for the leg animation, or apply a fixed animation. These solutions introduce highly noticeable foot sliding and look artificial. In contrast, our solution combining Motion Matching with orientation prediction produces natural-looking walking animations with smooth transitions between different velocities and orientations, thus better adjusting the avatar's movement to the user.

\paragraph*{Body and head orientations}
When using natural walk to navigate in immersive VR, it is essential to keep the head and torso orientation decoupled so that the user is free to move in any direction while rotating the HMD to look around. HMD velocity should also be distinguishable from torso movement so that the user can take steps in any direction: from walking forward to side-stepping. Most applications use the HMD's forward direction to orient the avatar's body or rotate the avatar's body when the angle between the HMD's direction and the current body direction is above a certain threshold. These solutions keep the avatar torso from correctly aligning with the user and often result in wrong motions when applying procedural animation. Our neural network predicts the user's body orientation from the trackers' data so that the body can be correctly oriented when combined with Motion Matching. 

\paragraph*{Non-upright motion}
Another essential aspect when animating full-body avatars is their behavior when users bend their legs. Those applications that allow the avatar to crouch typically require the user to press a button or use the HMD's height information to bend the legs procedurally. These approaches usually result in incorrect pose matching and unnatural poses. Our approach achieves non-upright motion (e.g., tip-toes or crouching) by changing the animation database, which can be switched at runtime based on a given parameter or condition.

\paragraph*{In-place rotations}
In-place rotations are another common movement in VR that the user often performs while looking around, and they typically require small steps for changing the body orientation. Most applications handle this user movement by rotating the avatar in place, keeping a static pose or by applying a slow walk forward animation, but both cases result in noticeable foot sliding. Motion Matching naturally handles changes in direction that require small steps since it is part of the trajectory features in the query vector, thus allowing us to replicate such behavior.

\subsection{Modifying the animation database}
Motion Matching has mainly been used for locomotion systems, but different types of motions can be represented by selecting different features. However, we can still achieve different behaviors without redefining the features; only by changing the animation database, we can automatically get significantly different animations. For instance, we could capture different walking gaits and get a completely different set of animations without changing the system.

Recently, a large number of dancing applications have been released for VR. Like in our work, inverse kinematics for the upper body is the most used solution. However, dragging the avatar or a fixed animation is usually used for the lower body. We used motion capture to record a user dancing and created an animation database. Only changing the database, we obtained high-quality dancing animations with transitions depending on the user's physical walking and body orientation. The result is shown in Figure~\ref{fig:res:dancing}.

\begin{figure}[h]
  \centering
  \includegraphics[width=1.0\linewidth]{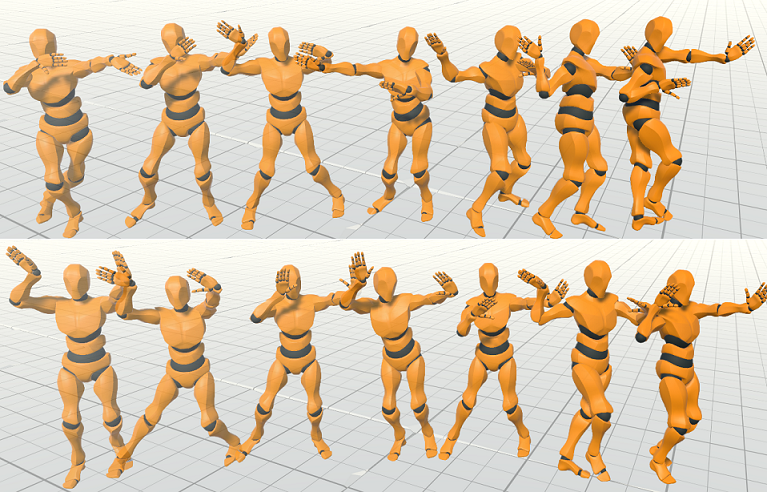}
  \caption[VR Avatar dancing]{\label{fig:res:dancing} VR avatar animated with Motion Matching and a dancing animation database.}
\end{figure}

\subsection{Animation database effect on the position accuracy} 
The positional accuracy of the search results is limited by the discrete number of animations in the database. In Section \ref{sec:vr:pos_accuracy} we explained the challenge of enforcing the position of the avatar to a specific location when animating self-avatars with Motion Matching, which is aggravated in VR due to the highly unpredictable nature of the user trajectory. Consequently, to keep the avatar at a reasonable distance from the user, we bound the positional error between the user and the avatar and clamp the avatar position if necessary (thus, introducing some foot sliding).

\begin{figure*}[h]
  \centering
  \includegraphics[width=1\linewidth]{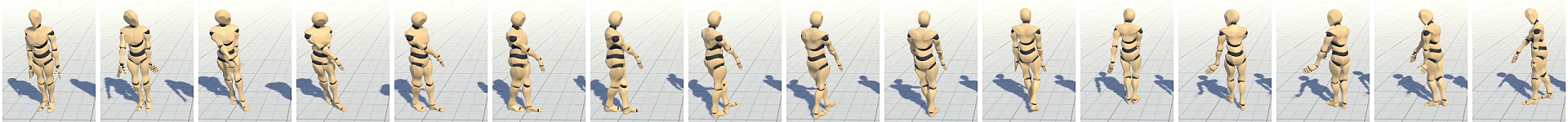}
  \includegraphics[width=1\linewidth]{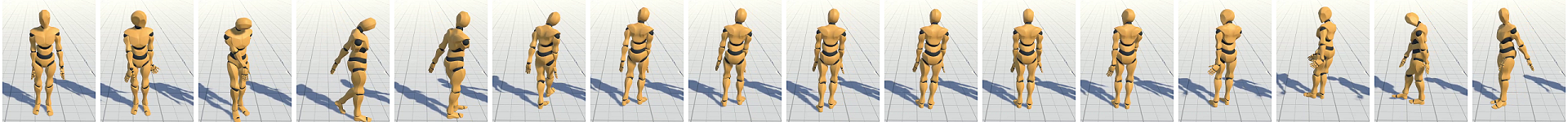}
  \includegraphics[width=1\linewidth]{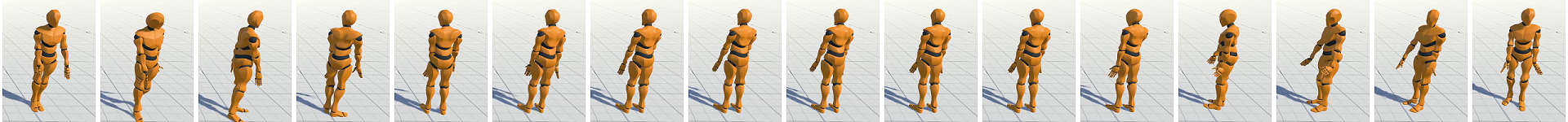}
  \caption[Animation database size comparison]{\label{fig:res:animationdb_comparison} Three avatars animated with the same input (user doing a turn-in-place), but using an increasingly large portion of the animation database. From top to bottom: 10\%, 25\% and 100\% of the poses in the database. Smaller databases struggle to match the user's motion, thus reducing the final quality and increasing the time and space needed to complete the turn. }
\end{figure*}

Further extending the animation database with a large variety of motions can reduce the positional error between the user and the avatar. Ideally, if the animation database could represent all possible user movements in VR, Motion Matching could always perfectly follow the user, and there would be no positional error. Therefore, there is a strong dependency between the positional accuracy of Motion Matching and the size and variety of the database. Figure \ref{fig:res:animationdb_comparison} shows the effect of applying different sizes of animation databases to the same user input to show the importance having a good database on the final quality of the movements and the positional accuracy.

\begin{figure}[h]
  \centering
  \includegraphics[width=0.6\linewidth]{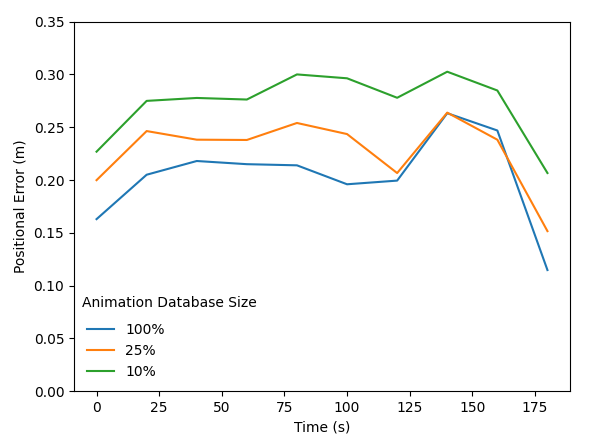}
  \caption[Position accuracy for different animation database sizes]{\label{fig:res:animationdb_positionalerror} Positional error (distance between the user and the avatar) for different animation database sizes using the same user input. 100\% is our complete animation database, while the others contain 25\% and 10\% of the poses in the complete database. The error is averaged every 20 seconds. Large databases represent a wider range of motions and better match the user's trajectories, thus minimizing the positional error.}
\end{figure}

We tested the system with our complete animation database and two reduced versions with 25\% and 10\% of the poses. The position accuracy $\alpha$ was set to 30\,cm. We performed different types of locomotion including run, walk, turn in place, walk in circles, and step sideways, while running Motion Matching for each database. For every frame, the positional error was computed as the distance between the target position (user) and the position of the virtual root joint (avatar). As shown in Figure \ref{fig:res:animationdb_positionalerror}, the positional error is reduced as the size of the database increases. Some movements had the same error for the complete database and the 25\% version due to the movement being well represented in both databases. The mean positional error for the complete database was 19\,cm, while for the 25\% and 10\% it was 22\,cm and 27\,cm, respectively.

\subsection{Position accuracy effect on the animation quality}

The position accuracy parameter $\alpha$ defined in Section \ref{sec:vr:pos_accuracy} is an upper bound of the positional error between the user and the avatar. On the one hand, this parameter should be as low as possible to maintain a good match between the user and the avatar positions. On the other hand, Motion Matching may need some flexibility to find a suitable trajectory to reach the target. The search may not find an exact match to the target trajectory at all times. It may sometimes deviate from the target, causing a positional error, but eventually, it will correct the deviation by searching for new trajectories towards the target. If a maximum positional error is enforced, it may be applied before Motion Matching can correct the trajectory and reduce the animation quality by limiting the number of poses used.

\begin{figure}[h]
  \centering
  \includegraphics[width=0.5\linewidth]{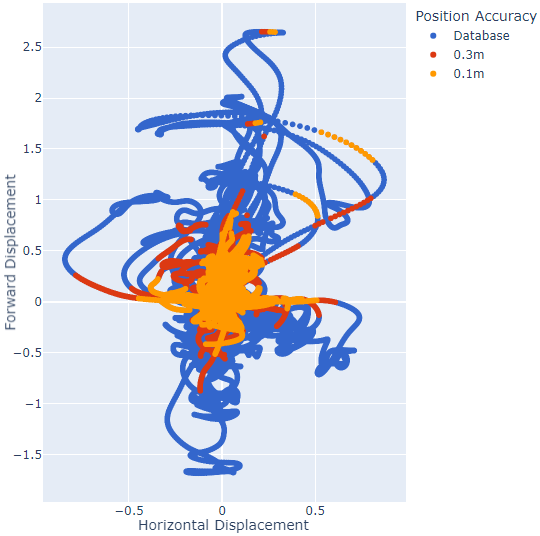}
  \caption[Trajectories depending on the position accuracy]{\label{fig:res:coverage} Poses used by two avatars with $\alpha=0.3$\,m and $\alpha=0.1$\,m were recorded for 5 minutes. The plot shows one of the 2D position features (0.33 seconds in the future) from a top-down view. Larger position accuracy allows Motion Matching to use more variety of poses.}
\end{figure}

We set up two avatars with $\alpha=0.3$\,m and $\alpha=0.1$\,m, and controlled them simultaneously for around 5 minutes. We recorded the indices to the pose and feature databases used for each avatar. Figure \ref{fig:res:coverage} shows one of the 2D position features (0.33 seconds in the future) for all poses in the database and for those poses used for the avatars. The avatar with $\alpha=0.3$\,m has more freedom of movement and can use a larger number of poses, thus, enhancing the final animation quality. In total, the avatar with $\alpha=0.3$\,m used 6,932 different poses while the other used 5,715 different poses. 

\subsection{Body orientation prediction}
In this section, we compare our neural network and the HMD's forward direction for predicting body orientation. We also compare the accuracy of the neural network depending on the parameter $r$ defined in Section~\ref{sec:vr:network_training}. Xsens was used to capture the motion of a user playing a video game for Oculus Quest for 15 minutes. The game required the user to change the direction of the body and the head frequently. We measured the angle error between the ground truth body direction captured with Xsens, which is the projection of the hip joint forward direction onto the ground, and the different body orientation predictors. 

\begin{figure}[h]
  \centering
  \includegraphics[width=0.6\linewidth]{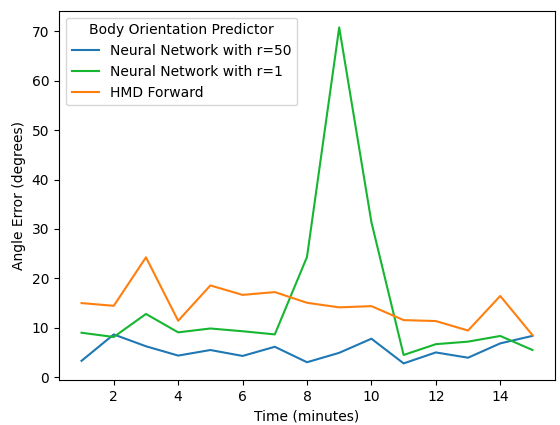}
  \caption[Body direction predictors comparison]{\label{fig:res:body_direction_predictors} Average angle error per minute for different body predictors while playing a video game for 15 minutes. HMD's forward direction had a larger error than our neural network for predicting the body orientation. The neural network trained with $r=1$ learns to imitate the previous orientation instead of using the trackers’ data resulting in a larger error than training with $r=50$.}
\end{figure}

Figure~\ref{fig:res:body_direction_predictors} shows the average angle error per minute for the different body predictors. Directly using the HMD's forward direction as body direction had a mean angle error of 14.5º and a standard deviation of 18.9º, while using our neural network trained with $r=50$ had a mean angle error of 5.4º and standard deviation of 7.7º. When $r=1$, the neural network learns to imitate the previous orientation, which is one of the network's inputs because it always comes from the ground truth data during training. At runtime, the previously predicted orientation is given as an input to the network, therefore, it may not be reliable. The mean angle error when $r=1$ is 15.0º, and the standard deviation is 25.7º. The result is worse than when using $r=50$ because, as shown in Figure~\ref{fig:res:body_direction_predictors}, around minute 7, the angle error is considerable, and since the neural network is imitating previously predicted orientations, it cannot quickly recover from the failure.

\chapter{Conclusions and Future Work} \label{chap:conclusions}

\section{Conclusions}
This work contributes to existing knowledge in data-driven methods for computer animation by providing a complete study and development of Motion Matching for character animation and VR avatars. We comprehensively explain all necessary components to implement Motion Matching: motion capture; animation, pose, and feature databases; character controllers; and blending inertialization. We also implement the system in the game engine Unity so that it can be reused in any other project, thus providing high-quality animations with minimal setup.

The use of Motion Matching for character animation has been widely adopted for complex tasks such as locomotion due to its versatility. Automatically, it can represent all types of transitions and motions without requiring complex state machines. As a result, it provides a robust, scalable framework to animate virtual characters in complex environments. 

This project also presents a novel method for using Motion Matching combined with Deep Learning to animate virtual reality avatars. By combining inverse kinematics with data-driven strategies, we can obtain natural behaviors while preserving accuracy in the upper-body poses. In addition, we propose a neural network for predicting the user's body orientation from the tracking data available in consumer-grade VR devices. We hope our work provides a foundation to continue researching data-driven methods for VR.

\section{Future work}
Although Motion Matching has successfully demonstrated its capacity and versatility to create high-fidelity animations, it is usually combined with traditional systems to work around its limitations. For instance, in this work, we combine it with IK to improve the position accuracy for upper-body poses. More research is required to understand data-driven methods better and overcome their limitations. Here, we suggest some lines of research:
\begin{itemize}
    \item Continuous motion manifold representation. One of the limitations of Motion Matching is the discrete nature of the animation database. By using Deep Learning-based methods, it may be possible to represent the set of human motion and use it to obtain a continuous representation. This approach would further create sequences of poses that accurately match the desired trajectories and reduce the overall memory requirements.
    \item Search approach. Currently, the search of poses is formulated as a greedy nearest-neighbor search. No evaluation of the overall quality of the animations from a long-term perspective is performed. Reinforcement Learning algorithms may help find a better policy for selecting subsequent poses.
    \item Automatic feature extraction. One of the most challenging tasks when using Motion Matching is selecting the appropriate features to represent the desired motions. Other fields such as Computer Vision already perform automatic extraction of features; this idea could be researched for the field of computer animation to bring data-driven methods to a broader range of motions, such as for full-body VR avatars.
\end{itemize}

\printbibliography[heading=bibintoc, title={Bibliography}]

\end{document}